\documentclass{aa}
\usepackage{graphicx}
\usepackage{psfig}

\newcommand{\gtrsim}{\mathrel{\hbox{\rlap{\lower.55ex \hbox {$\sim$}}
                   \kern-.3em \raise.4ex \hbox{$>$}}}}
\newcommand{\lesssim}{\mathrel{\hbox{\rlap{\lower.55ex \hbox {$\sim$}}
                   \kern-.3em \raise.4ex \hbox{$<$}}}}

\begin{document}

   \title{Photospheric radius expansion X-ray bursts as standard candles}

   \author{E.~Kuulkers
           \inst{1,2,}
\thanks{\emph{Present address:} ESA-ESTEC,
  Science Operations \&\ Data Systems Division,
  SCI-SDG,
  Keplerlaan 1,
  2201 AZ Noordwijk, The Netherlands}
	   \and
	   P.R.~den Hartog
	   \inst{1,2}
	   \and
	   J.J.M.~in 't Zand
	   \inst{1}
	   \and
	   F.W.M.~Verbunt
           \inst{2}
	   \and
	   W.E.~Harris
	   \inst{3}
	   \and
	   M.~Cocchi
	   \inst{4}
          }

   \offprints{Erik Kuulkers}

   \institute{
	      SRON National Institute for Space Research,
	      Sorbonnelaan 2, 3584 CA Utrecht, The Netherlands\\
              \email{Erik.Kuulkers@rssd.esa.int}
	      \and
	      Astronomical Institute, Utrecht University,
	      P.O.\ Box 80000, 3508 TA Utrecht, The Netherlands
	      \and
	      Department of Physics and Astronomy, 
	      McMaster University, Hamilton, ON L8S 4M1, Canada
	      \and
	      Istituto di Astrofisica Spaziale e Fisica Cosmica (IASF)
	      Area Ricerca Roma Tor Vergata, 
	      Via del Fosso del Cavaliere,
	      I-00133 Roma, Italy
             }

   \date{Received --; accepted --}

   \titlerunning{Photospheric radius expansion X-ray bursts as standard candles}

\abstract{
We examined the maximum bolometric peak luminosities during 
type~I X-ray bursts from the persistent or transient luminous X-ray sources in globular clusters. 
We show that for about two thirds of the sources 
the maximum peak luminosities during photospheric radius expansion 
X-ray bursts extend to a critical value of 3.79$\pm$0.15$\times$10$^{38}$\,erg\,s$^{-1}$, 
assuming the total X-ray burst emission is entirely due to black-body radiation and the 
recorded maximum luminosity is the actual peak luminosity.
This {\it empirical} critical luminosity is consistent with the Eddington 
luminosity limit for hydrogen poor material. Since the critical luminosity
is more or less always reached during photospheric radius expansion X-ray
bursts (except for one source), such bursts may be regarded as empirical standard candles.
However, because significant deviations do occur, our standard candle is only accurate 
to within 15\%.
We re-evaluated the distances to the twelve globular clusters in which the 
X-ray bursters reside.
  }

   \maketitle

\section{Introduction}

\subsection {Type~I X-ray bursts}

Type~I X-ray bursts (hereafter X-ray bursts, except when needed) in low-mass X-ray binaries 
are thermonuclear runaways in freshly accreted material on the surface of a neutron star
(for a review see Lewin et al.\ 1993). 
The freshly accreted matter is compressed and heated, on a time scale of hours to days, to densities
and temperatures adequate for thermonuclear ignition.
Due to the strong temperature sensitivity of the nuclear
reactions, the ignition nearly always results in a runaway process
that leads to a sudden, rapid (between about 1 to 10\,sec) increase in the X-ray
flux. These X-ray bursts last generally for tens to hundreds of seconds, and recur
with a frequency (partly) set by the supply rate of fresh fuel.  The
spectral hardening during the X-ray burst rise and subsequent softening
during the decay reflect the heating 
and subsequent cooling of the neutron star surface. The X-ray spectra during the X-ray bursts 
are consistent with black-body emission from a compact object with a radius of approximately 10\,km 
and temperature between 1 and 2.5\,keV.

Spikes with similar duration in X-ray light curves have been seen
in MXB\,1730$-$335, also known as `The Rapid Burster' (e.g.\ Lewin et al.\ 1993, and references therein; this
source also shows type~I X-ray bursts),
and GRO\,J1744$-$28, or `The Bursting Pulsar' (Lewin et al.\ 1996; Kommers et al.\ 1997),
but these do not show the characteristics of type~I X-ray bursts (in particular, the cooling during the decay).
They are thought to be due to sudden accretion events and are referred to as type~II X-ray bursts 
(Hoffman et al.\ 1978).

During some type~I X-ray bursts the energy release is high
enough that the luminosity at the surface of the neutron
star reaches the Eddington limit. At that point the neutron star atmosphere expands due to
radiation pressure. During expansion and subsequent contraction the luminosity is expected to
remain almost constant near the Eddington limit. Since the luminosity scales as $L_{\rm b}\propto R^2T^4$
for pure black-body radiation, the effective temperature, $T$, drops
when the radius of the photosphere, $R$, expands.
X-ray bursts with a photospheric radius expansion/contraction phase are therefore recognized by
an increase in the inferred radius with a simultaneous decrease in the effective temperature near the
peak of an X-ray burst, at approximately constant
observed flux. 
The point at which the neutron star atmosphere reaches the surface again (i.e.\ at the highest effective temperatures)
is called `touch-down'. When the expansion is large the
temperature may become so low that the peak of the radiation shifts to the UV wavelengths,
and no or little X-rays are emitted. Such photospheric radius expansion events
(hereafter radius expansion bursts, except when needed) are recognizable by
so-called `precursors' in the X-ray light curves followed by a `main' burst (Tawara et al.\ 1984; Lewin et al.\ 1984).
 
We note that, although the X-ray burst spectra seem to be adequately described by black-body emission, 
the `true' emission from the neutron star and its close environment is expected to be more complex.
Electron scattering dominates the opacity in the neutron star atmosphere,
leading to deviations from the original Planckian spectrum (van Paradijs 1982;
London et al.\ 1984, 1986; see also Titarchuk 1994; Madej 1997, and references therein).
Similarly, electron scattering in a disk corona (e.g.\ Melia 1987) and/or winds or outflows
from the neutron star surface interacting with the accretion disk (e.g.\ Melia \&\ Joss 1985;
Stollman \&\ van Paradijs 1985) can affect the original X-ray spectrum.

\subsection{A standard candle?}

Shortly after the discovery of X-ray bursts, van Paradijs (1978) introduced the idea that 
the average peak luminosity of type I X-ray bursts is a standard candle, presumably the Eddington luminosity limit
for the neutron star photosphere. 
With that assumption, approximate neutron star radii and 
distances could be derived for a sample of X-ray bursters. 
Scaling the derived distances so that they are symmetrically distributed around the 
Galactic center at $\simeq$9\,kpc, van Paradijs (1981) found the average peak (bolometric) luminosity
to be 3$\times$10$^{38}$\,erg\,s$^{-1}$. 
Van Paradijs (1981) realized that X-ray bursters residing in globular clusters
are important calibrators, since their distances can be determined independently.
Their peak luminosity agreed with the above quoted value of the average peak luminosity of 
Galactic center X-ray bursters. Verbunt et al.\ (1984) redid this work, based upon a more extensive
sample of X-ray bursters. They found the average peak luminosity of X-ray bursters not located 
in globular clusters to be 3.5$\pm$1.0$\times$10$^{38}$\,erg\,s$^{-1}$, when taking for 
the distance to the Galactic center a value of 8.5\,kpc. For the X-ray bursters in globular clusters a 
luminosity value of 4.0$\pm$0.9$\times$10$^{38}$\,erg\,s$^{-1}$
was derived, i.e.\ consistent with the above value. A `standard candle' value of 
3.7$\times$10$^{38}$\,erg\,s$^{-1}$ was adopted.
Cominsky (1981), on the other hand, assumed that the black-body radius is 
the same for all X-ray bursts and subsequently derived source distances and X-ray burst peak luminosities.
Again scaling the average source distance to 9\,kpc, peak luminosities
between $\simeq$10$^{38}$ and $\simeq$5$\times$10$^{38}$\,erg\,s$^{-1}$ were found.

Van Paradijs (1978, 1981) cautioned, however, that in a few individual sources the scatter in the 
X-ray burst peak luminosity was appreciable. It was therefore suggested by Lewin (1982) to use instead 
only the brightest X-ray bursts. In this way the standard candle became $\gtrsim$5.5$\times$10$^{38}$\,erg\,s$^{-1}$.
Of course, the problem arises that one does not know whether
the strongest X-ray bursts observed are also the strongest ones possible. It seemed, however, that 
those X-ray bursts which showed clear photospheric radius expansion do reach a `true' critical
luminosity. 
But whether this critical luminosity is also a standard candle was still questioned
(Basinska et al.\ 1984). If that would turn out to be the case one can in principle determine
the distances to sources which show clear radius expansion bursts; for those sources
whose X-ray bursts do not show a radius expansion phase we can still determine an upper limit to the 
distance using the maximum observed peak flux.

The thirteen persistent or transient 
luminous ($\gtrsim$10$^{36}$\,erg\,s$^{-1}$) X-ray sources in twelve globular clusters
are all low-mass X-ray binaries
(e.g.\ Verbunt et al.\ 1984; Hut et al.\ 1992; White \&\ Angelini 2001; in 't Zand et al.\ 2001b). 
All but one (AC\,211 in NGC\,7078) have shown type~I X-ray bursts\footnote{During a recent
outburst GRS\,1747$-$312 in Terzan 6 showed type~I X-ray bursts (in~'t~Zand et al.\ 2002, in preparation).}, 
proving they harbour neutron stars. Several of these X-ray bursts exhibited a photospheric radius expansion phase 
(see, e.g., Lewin et al.\ 1993, and references therein). 
We here show that the peak luminosity of radius expansion bursts 
from about two third of the X-ray bursters in globular clusters do indeed reach 
an empirical critical luminosity, and may be used 
as approximate standard candles for other sources.

\section{Distances to 12 globular clusters}

\begin{table*}
\caption{Parameters per globular cluster needed for evaluating their distances, $d$.
References:
[1] Harris (1996),
[2] Pritzl et al.\ (2001),
[3] Heasley et al.\ (2000),
[4] Chaboyer et al.\ (2000),
[5] Paltrinieri et al.\ (2001),
[6] Idiart et al.\ (2002),
[7] Cohn et al.\ (2002),
[8] Davidge (2000),
[9] Origlia et al.\ (2002).
}
\begin{tabular}{ccccccc}
\hline
\multicolumn{7}{c}{~} \\
Cluster  & $E(B-V)$    &  $V_{\rm HB}$      &   [Fe/H]    &  ref  &  $(m-M)_0$  &  $d$ (kpc) \\
\multicolumn{7}{c}{~} \\
\hline
\multicolumn{7}{c}{~} \\
NGC\,1851 & 0.02 $\pm$ 0.01 & 16.09 $\pm$ 0.05 & $-$1.22 $\pm$ 0.1  & [1]     & 15.41 $\pm$ 0.06 & 12.1 $\pm$ 0.3 \\
NGC\,6440 & 1.07 $\pm$ 0.10 & 18.70 $\pm$ 0.20 & $-$0.34 $\pm$ 0.2  & [1]     & 14.63 $\pm$ 0.36 & 8.4 $^{+1.5}_{-1.3}$ \\
NGC\,6441 & 0.51 $\pm$ 0.05 & 17.51 $\pm$ 0.05 & $-$0.53 $\pm$ 0.2  & [1],[2] & 15.21 $\pm$ 0.17 & 11.0 $^{+0.9}_{-0.8}$ \\
NGC\,6624 & 0.32 $\pm$ 0.03 & 16.10 $\pm$ 0.05 & $-$0.63 $\pm$ 0.1  & [3]     & 14.40 $\pm$ 0.11 & 7.6 $\pm$ 0.4  \\
NGC\,6652 & 0.12 $\pm$ 0.02 & 15.96 $\pm$ 0.05 & $-$0.90 $\pm$ 0.1  & [1],[4] & 14.92 $\pm$ 0.08 & 9.6 $\pm$ 0.4  \\
NGC\,6712 & 0.33 $\pm$ 0.05 & 16.25 $\pm$ 0.05 & $-$0.90 $\pm$ 0.1  & [1],[5] & 14.56 $\pm$ 0.16 & 8.2 $\pm$ 0.6  \\
NGC\,7078 & 0.10 $\pm$ 0.02 & 15.83 $\pm$ 0.05 & $-$2.25 $\pm$ 0.05 & [1]     & 15.06 $\pm$ 0.08 & 10.3 $\pm$ 0.4 \\
Terzan~1  & 2.00 $\pm$ 0.30 & 19.95 $\pm$ 0.20 & $-$1.3 $\pm$ 0.2   & [1],[6] & 13.14 $\pm$ 0.95 & 4.3 $^{+2.3}_{-1.5}$ \\
Terzan~2  & 1.57 $\pm$ 0.15 & 20.30 $\pm$ 0.20 & $-$0.40 $\pm$ 0.2  & [1]     & 14.88 $\pm$ 0.51 & 9.5 $^{+2.5}_{-2.0}$ \\
Terzan~5  & 2.18 $\pm$ 0.20 & 22.26 $\pm$ 0.30 &  0.00 $\pm$ 0.2  & [1],[7] & 14.70 $\pm$ 0.69 & 8.7 $^{+3.3}_{-2.4}$ \\
Terzan~6  & 2.14 $\pm$ 0.20 & 22.25 $\pm$ 0.20 & $-$0.50 $\pm$ 0.2  & [1]     & 14.89 $\pm$ 0.65 & 9.5 $^{+3.3}_{-2.5}$ \\
Liller~1  & 3.13 $\pm$ 0.20 & 25.20 $\pm$ 0.30 & $-$0.20 $\pm$ 0.2  & [8],[9] & 14.73 $\pm$ 0.69 & 8.8 $^{+3.3}_{-2.4}$ \\
\multicolumn{7}{c}{~} \\
\hline
\end{tabular}
\label{Harris}
\end{table*}

The distances to the twelve globular clusters
have been re-evaluated for the purposes of this
paper, to ensure that they are on an internally
homogeneous system.  We adopt the luminosity
level of the horizontal branch (or equivalently
the RR Lyrae stars) in the cluster as the fiducial
distance indicator, following Harris (1996).

The three key observational quantities needed
to deduce the distance are then the apparent magnitude
of the horizontal branch, $V_{\rm HB}$, the foreground
reddening, $E_{B-V}$, and the metallicity, [Fe/H].
The best recent estimates for these are listed in Table~\ref{Harris},
either from the catalog of 
Harris (1996)\footnote{The online catalogue of Milky Way globular 
cluster parameters was last revised on 1999, June 22
(see {\tt http://physun.physics.mcmaster.ca/Globular.html}).} 
or from the more recent literature as listed individually in the table.

The absolute calibration of distance follow the
prescription adopted by Harris (1996), namely
\begin{equation}
(m-M)_0 = V_{\rm HB} - M_V({\rm HB}) - 3.1~E(B-V) 
\end{equation}
where the absolute magnitude is a function of metallicity,
\begin{equation}
M_V({\rm HB}) = 0.15 {\rm [Fe/H]} + 0.80 \, .
\end{equation}
We note that $V_{\rm HB}$ and $M_V({\rm HB})$ refer to the
observed mean level of the horizontal branch
stars, and not the (slightly fainter) level of the
zero-age horizontal branch, $M_V({\rm ZAHB})$.  
The uncertainties in $V_{\rm HB}$, $M_V$, and $E(B-V)$ have
been added in quadrature to get the final uncertainty in 
the distance modulus, $(m-M)_0$.
The dominant source of uncertainty is either
the $V_{\rm HB}$ estimate or, for highly-reddened globular clusters, the 
uncertainty in the reddening.  In some cases the errors are
rather large; the last five clusters in Table~\ref{Harris} have formal
distance uncertainties that are on the order of 30\%.

In some other recent papers, a slightly steeper slope for
the relation between $M_V({\rm HB})$ and metallicity [Fe/H]
is used (see, e.g., Chaboyer et al.\ 1998, where 
$M_V({\rm HB}) = 0.23$ [Fe/H] + 0.83 is used).  These relations
are negligibly different from Eq.~(2) for 
the range [Fe/H] $> -1$ which covers most of our clusters,
and at worst gives a difference of $\simeq$0.1 magnitude
in the distance modulus for the most metal-poor cluster
in the list, a level which is within the current uncertainties
in the absolute distance scale for globular clusters.

It is worth noting that the results for the distances
in Table~\ref{Harris} ($d$-values in the last column) are consistent with
the view that all these clusters, except for NGC\,1851, NGC\,6441 and
NGC\,7078, are within the Galactic bulge:  they lie
roughly in the direction of the Galactic center, and have distances
in the approximate range 8$\pm$2\,kpc which place them in the 
extended bulge region.  Thus the very rough assumption that
$d \sim 8$\,kpc for all the clusters in our list, independent of
other observational information about their individual distances,
would not give a calibration of their X-ray burst luminosities
that was grossly in error.

\begin{figure}
\begin{center}
 \resizebox{\hsize}{!}{\includegraphics[angle=-90, clip, bb=190 52 582 475]{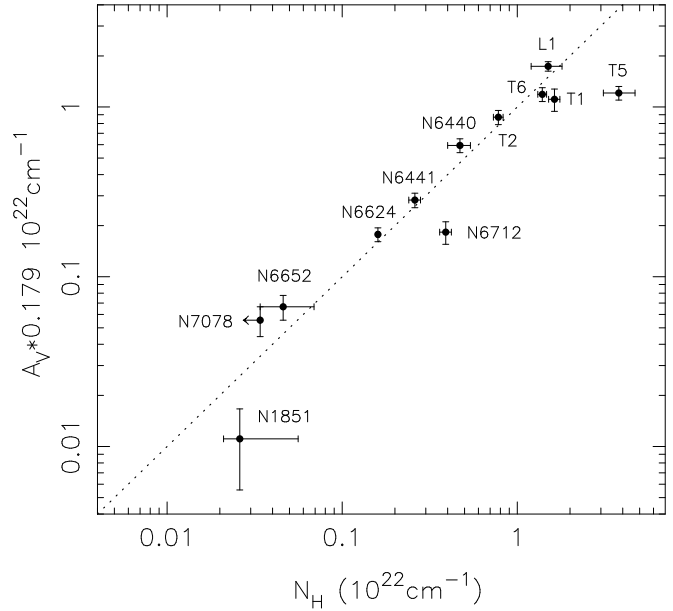}}
 \caption{Interstellar absorption as derived from the known relation between 
 optical extinction and dust (and hydrogen) column densities (Predehl \&\ Schmitt 1995)
 versus the interstellar absorption as derived from X-ray spectral fits, see text. We have indicated the different
 globular clusters as follows, e.g.: N1851 = NGC\,1851, T1 = 
Terzan~1, and L1 = Liller~1.
 \label{plot_nh}
}
\end{center}
\end{figure}

A strong correlation exists between the visual extinction, $A_V$, and the dust (and hydrogen) column densities,
if there is no additional absorption within the binary itself:
$N_{\rm H}$\,[cm$^{-2}$]/$A_V$=1.79$\times$10$^{21}$ (Predehl \&\ Schmitt 1995). In Fig.~\ref{plot_nh} we show
$N_{\rm H}$ as derived from the optical extinction towards the globular clusters in 
Table~\ref{Harris} ($A_V = 3.1 E(B-V)$) using the above relation, versus $N_{\rm H}$ as derived from the X-ray spectral fits
to the globular cluster source spectra, see Table~\ref{GC} (see also Sidoli et al.\ 2001). 
Most lie close to the correlation, except A1850$-$08/NGC\,6712,
XB\,1733$-$30/Terzan~1, and XB\,1745$-$25/Terzan~5. For these latter sources the discrepancy may 
be partly due to the X-ray continuum model used (XB\,1733$-$30; see also Sidoli et al.\ 2001, Table~\ref{tabter5}), 
energy range used in the X-ray spectral fitting (XB\,1745$-$25), or to intrinsic absorption near or in 
the binaries themselves.

\begin{table*}
\caption{
Parameters for the twelve X-ray bursters in globular clusters ordered along
right ascension. Given are the X-ray burst source name,
the globular cluster to which it belongs, 
the interstellar absorption column towards the source ($N_{\rm H}$), the
orbital period, $P_{\rm orb}$, if known, 
its bolometric black-body peak flux for radius expansion bursts, 
$\overline{F_{\rm bb,peak}}$, or the maximum bolometric black-body peak flux for normal
X-ray bursts, $F_{\rm bb,max}$ (see Appendix~\ref{account} for details), 
and whether it exhibited photospheric radius expansion (RE) X-ray bursts or not. 
References are given between brackets: 
[1] Sidoli et al.\ (2001), 
[2] Masetti et al.\ (2000),
[3] Parmar et al.\ (1989),
[4] this paper (see Appendix~\ref{ter5}),
[5] Parmar et al.\ (2001),
[6] White \&\ Angelini (2001),
[7] Deutsch et al.\ (2000), 
[8] Homer et al.\ (2001). 
[9] Sansom et al.\ (1993) and references therein,
[10] in~'t~Zand et al.\ (2000), 
[11] Stella et al.\ (1987), 
[12] Chou \&\ Grindlay (2001) and references therein,
[13] Heinke et al.\ (2001),
[14] Homer et al.\ (1996).} 
\begin{tabular}{cccccccc}
\hline
\multicolumn{8}{c}{~} \\
\multicolumn{1}{c}{X-ray burster} & \multicolumn{1}{c}{globular} & 
\multicolumn{1}{c}{$N_{\rm H}$} & \multicolumn{1}{c}{ref} &
\multicolumn{1}{c}{$P_{\rm orb}$} & \multicolumn{1}{c}{ref} &
\multicolumn{1}{c}{$\overline{F_{\rm bb,peak}}$ or $F_{\rm bb,max}$} & \multicolumn{1}{c}{RE?} \\
\multicolumn{1}{c}{~} & \multicolumn{1}{c}{cluster} & 
\multicolumn{1}{c}{(10$^{22}$\,cm$^{-2}$)} & \multicolumn{1}{c}{~} &
\multicolumn{1}{c}{(hr)} & \multicolumn{1}{c}{~} & 
\multicolumn{1}{c}{(10$^{-8}$\,erg\,s$^{-1}$)} & \multicolumn{1}{c}{~} \\
\multicolumn{8}{c}{~} \\
\hline
\multicolumn{8}{c}{~} \\
MX\,0513$-$40    & NGC 1851  & 0.026 $^{+0.030}_{-0.005}$ & [1] & $<$1? & [7],[8] & 2.00 $\pm$ 0.18 & yes \\
4U\,1722$-$30    & Terzan 2  & 0.78 $\pm $0.05 & [1] & & & 6.15 $\pm$ 0.09 & yes \\
MXB\,1730$-$335$^a$& Liller 1  & 1.5 $\pm$ 0.3 & [2] & & & 1.66 $\pm$ 0.08 & no \\
XB\,1733$-$30    & Terzan 1  & 1.63 $\pm$ 0.12 & [3] & & & 7.4  $\pm$ 1.0  & no \\
XB\,1745$-$25    & Terzan 5  & 3.8 $^{+0.9}_{-0.7}$ & [4] & & & 4.75 $\pm$ 0.12 & yes \\
MX\,1746$-$20    & NGC 6440  & 0.47 $\pm$ 0.07 & [1] & & & 1.77 $\pm$ 0.14 & no \\
4U\,1746$-$37    & NGC 6441  & 0.26 $\pm$ 0.02 & [1] & 5.7 & [9] & 0.95 $\pm$ 0.07 & yes \\
GRS\,1747$-$312   & Terzan 6  & 1.39 $\pm$ 0.08 & [1] & 12.4 & [10] & 1.71 $\pm$ 0.06 & yes \\
4U\,1820$-$30    & NGC 6624  & 0.16 $\pm$ 0.003 & [1] & 0.19 & [11],[12] & 5.27 $\pm$ 0.72  & yes \\
H1825$-$331    & NGC 6652  & 0.046 $^{+0.023}_{-0.012}$ & [5] & 0.92 or 2.2\,hr? & [13] & 2.87 $\pm$ 0.07 & no \\
A1850$-$08    & NGC 6712  & 0.39 $^{+0.01}_{-0.03}$ & [1] & 0.34 & [14] & 5.2 $\pm$ 0.5 & yes \\
4U\,2129+12$^b$  & NGC 7078  & $<$0.034 & [6] & & & 3.94 $\pm$ 0.29 & yes \\
\multicolumn{8}{c}{~} \\
\hline
\multicolumn{8}{l}{\footnotesize $^a$\,More popularly known as the Rapid Burster.} \\
\multicolumn{8}{l}{\footnotesize $^b$\,NGC\,7078 X-2 is the likely source of the X-ray bursts (White \&\ Angelini 2001).} \\
\end{tabular}
\label{GC}
\end{table*}

\section{Observations and analysis}

For our investigation we analysed data obtained by the Proportional Counter Array (PCA) onboard
{\em Rossi X-ray Timing Explorer} (RXTE) and the Wide Field Cameras (WFCs) onboard the 
{\em `Satellite per Astronomia X'} (BeppoSAX).
RXTE and BeppoSAX have been operational for more than five years, and have
observed most of the bright globular cluster X-ray sources. 
We complement these with results reported in the literature.

\subsection{RXTE/PCA}

The PCA
(2--60\,keV; Bradt et al.\ 1993) combines a large collecting area (maximum of $\simeq$6500\,cm$^2$) 
with high time resolution (down to $\mu$s). This is ideal
to study spectral variations on short time scales, such as which occur during X-ray bursts. 
We inspected light curves from the publicly available RXTE 
database of globular cluster X-ray bursters (up to January 2002, using the 
RXTE Master Catalog) at 1\,s time resolution using
the data collected in the {\sc Standard}~1 mode.
If more than one X-ray burst from an individual source was found, 
we analysed the bright X-ray bursts (which presumably reach the highest luminosity and may exhibit photospheric radius
expansion). 

We (re)analysed the X-ray bursts seen by the RXTE/PCA, all in a similar way, using the
latest information available on the response of the instrument at the relevant times.
Data were recorded in various modes (apart from the two standard modes) at high time resolution.
We used the data from the Burst Catcher (2\,ms resolution; 4U\,1722$-$30), 
Event (either 16, 125 or 500\,$\mu$s; MX\,0513$-$40, MXB\,1730$-$335, 4U\,1820$-$30, 4U\,2129+12),
or GoodXenon (4U\,1746$-$37, H1825$-$331) modes. The two former modes provide data in 64 channels 
covering the whole PCA energy range, while the latter provide data in all 256 channels available. 
The GoodXenon modes data were stored for every 2\,s; the high count rates reached during maximum 
of the X-ray bursts from 4U\,1746$-$37 and H1825$-$331 therefore resulted in data losses near the end 
of the 2\,s buffers. Spectra accumulated during these data losses were not taken into account. 
During the X-ray bursts reported here all Proportional Counter Units (PCUs) were on, 
except for MX\,0513$-$40 (only PCUs 0,1,2,3 on) and 4U\,2129+12 (only PCUs 0,2,3 on).
We created time resolved spectra at 0.25\,s resolution throughout the whole X-ray burst.
For a uniform analysis and in order to have as many photons as possible we used information 
from all PCU layers and as many PCUs as possible.
The count rate spectra were dead-time corrected and
a 1\%\ systematic uncertainty was included. 
The X-ray bursts seen from GRS\,1747$-$312 (in 't Zand et al.\ 2002, in preparation) were 
analysed in a similar way as the other X-ray bursts seen with the RXTE/PCA described here. 
During the observation of one of these X-ray bursts, used in our sample, only PCUs 0,2,3 were on. 

\begin{table*}
\caption{Spectral fit results for Crab (pulsar plus nebula) observations, using an absorbed
power law (Eq.~(2)), for various instruments (for the RXTE/PCA we also give the values for different
PCUs combined). Errors on the fit parameters were not always given by other authors; 
also sometimes $\Gamma$ and $N_H$ were fixed to the `canonical' values.
Although quoted by various authors, we note that the value of $N_H$=3$\times$10$^{21}$\,cm$^{-1}$ for the 
`canonical' spectrum is not derived by Toor \&\ Seward (1974) themselves.
Toor \&\ Seward (1974)  list various
estimates of $N_H$ derived by other authors; the average (no weighting) and the weighted average
of these estimates are 2.8$\pm$0.5 and 2.3$\pm$0.2$\times$10$^{21}$\,cm$^{-1}$, respectively.
The references are:
[1] Toor \&\ Seward (1984), 
[2] this paper (Sect.~3.5),
[3] Wilms et al.\ (1999), 
[4] Barret al.\ (2000),
[5] Christian \&\ Swank (1997),
[6] Parmar \&\ Smith (1985),
[7] Turner et al.\ (1989),
[8] in 't Zand (1999),
[9] Cusumano, G.\ (2002, private communication).
The last column gives the observed X-ray flux, $F_X$, in 10$^{-8}$\,erg\,cm$^{-2}$\,s$^{-1}$
in the 3--20\,keV energy band, uncorrected for interstellar absorption.}
\begin{tabular}{lcccccc}
\hline
\multicolumn{7}{c}{~} \\
instrument    & $\Gamma$ & $A$  & $N_H$& comments$^a$  & ref & $F_X$\\
\multicolumn{7}{c}{~} \\
\hline
\multicolumn{7}{c}{~} \\
`canonical'   & 2.10$\pm$0.03   & 9.7$\pm$1.0  & 3             &               & [1] & 2.38 \\
\multicolumn{7}{c}{~} \\
RXTE/PCA      & & & & & \\
PCUs 0,1,2,3 & 2.119$\pm$0.003 & 12.2$\pm$0.1 & 3 & 7.10 & [2] & 2.89 \\
PCUs 0,2,3   & 2.104$\pm$0.003 & 12.0$\pm$0.1 & 3 & 7.10 & [2] & 2.93 \\
all PCUs     & 2.130$\pm$0.003 & 12.8$\pm$0.1 & 3 & 7.10 & [2] & 2.96 \\
all PCUs     & 2.187 & 13.3    & 2.54 & 2.2.1            & [3] & 2.75 \\
all PCUs     & 2.18  & 13.5    & 3.3  & 2.36             & [4] & 2.82 \\
\multicolumn{7}{c}{~} \\
Einstein/MPC  & 2.1             & 9.5          & 3.45$^b$      &               & [5] & 2.13 \\
EXOSAT/ME     & 2.1             & 9.53         & 3             &               & [6] & 2.16 \\
Ginga/LAC     & 2.08$\pm$0.03   & 9.15$\pm$0.5 & 3             &               & [7] & 2.17 \\
BeppoSAX/WFC \#1& 2.1             & 9.77$\pm$0.26& 3$^c$       &               & [8] & 2.31 \\
BeppoSAX/WFC \#2& 2.1             & 9.93$\pm$0.05& 3$^c$       &               & [8] & 2.35 \\
BeppoSAX/MECS$^d$ & 2.093$\pm$0.001 & 9.15$\pm$0.01& 3.34$\pm$0.04 & Nov 1998  & [9] & 2.09 \\
\multicolumn{7}{c}{~} \\
\hline
\multicolumn{7}{l}{\footnotesize $^a$\,Version of response matrices.} \\
\multicolumn{7}{l}{\footnotesize $^b$\,Value taken from Schattenburg \&\ Canizares (1986).} \\
\multicolumn{7}{l}{\footnotesize $^c$\,Value of $N_{\rm H}$ was not quoted by the author. We assume here $N_{\rm H}$=3$\times$10$^{21}$\,cm$^{-1}$.} \\
\multicolumn{7}{l}{\footnotesize $^d$\,Spectral fits to Crab observations taken in 2000 and 2001.} \\
\end{tabular}
\label{Crab}
\end{table*}

\subsection{BeppoSAX/WFCs}

The WFCs onboard BeppoSAX are two coded aperture cameras 
(Jager et al.\ 1997). They point in opposite directions of one another and perpendicular to the 
Narrow-Field Instruments on the same satellite. The field of view of each WFC 
is 40 by 40 square degrees full-width to zero response with an angular resolution of about 5 arcmin in each 
direction. It is ideal to catch fast (seconds to minutes) phenomena at (un)expected places in the X-ray sky,
such as X-ray bursts.
Although it has a much lower collecting area (factor of $\sim$50 less than the RXTE/PCA), 
it has a much larger field of view 
(factor of $\sim$40 larger than the RXTE/PCA). Moreover, 
a large amount of time is devoted to the Galactic Center region (see, e.g., in~'t~Zand 2001),
so that a considerable part of the 
X-ray burster population is being monitored. 
In this paper we use those type~I X-ray bursts from globular clusters which showed either 
long enough expansion/contraction phases (typically $>$10\,s, see e.g., Cocchi et al.\ 2001), 
or X-ray bursts which had a short ($\sim$sec) expansion/contraction phase but which were bright, such as
those from 4U\,1820$-$30,
so that we have enough signal to noise for a meaningful X-ray spectral analysis during that phase.

\begin{figure*}
\begin{center}
 \resizebox{17cm}{!}{\includegraphics[angle=-90, clip, bb=31 37 563 739]{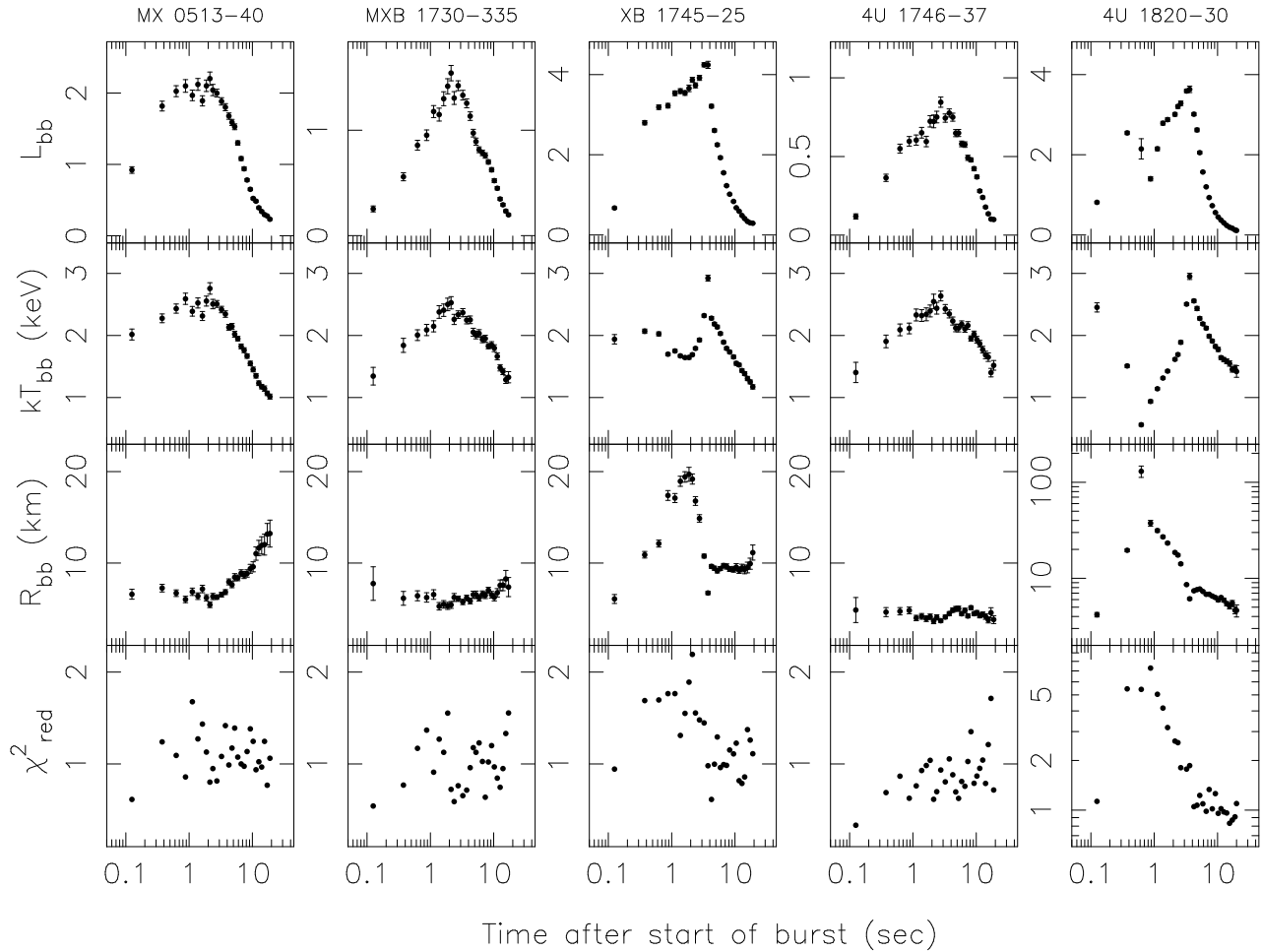}}
 \caption{Spectral fit results for the net X-ray burst emission of short ($\lesssim$30\,s) X-ray bursts from five globular cluster
 X-ray sources seen by the RXTE/PCA. The values are plotted on a logarithmic time scale; 
 they have also been logarithmically rebinned in time for clarity.
 From top to bottom: bolometric black-body luminosity, L$_{\rm bb}$, in 10$^{38}$\,erg\,s$^{-1}$,
 black-body temperature, kT$_{\rm bb}$, apparent black-body radius, R$_{\rm bb}$,
 and goodness of fit expressed in reduced $\chi^2$, $\chi^2_{\rm red}$. L$_{\rm bb}$ and R$_{\rm bb}$ are derived
 using the distances quoted in Table 1. Note that only the temperature axis has been kept the same
 for all the sources, and that the R$_{\rm bb}$ and $\chi^2_{\rm red}$ axes for 
 4U\,1820$-$30 have a logarithmic scale.
 T=0\,s corresponds to UT 1999 May 26 06:53:08, 1998 February 19 14:29:51, 2000 October 2 15:18:55, 1996 October 27 09:01:51 
 and 1997 May 2 17:32:45, for MX\,0513$-$40, MXB\,1730$-$335, XB\,1745$-$25, 4U\,1746$-$37 and 4U\,1820$-$30,
 respectively.
 \label{plot_5RXTEbursts}
}
\end{center}
\end{figure*}

\begin{figure*}
\begin{center}
 \resizebox{12cm}{!}{\includegraphics[angle=-90, clip, bb=31 37 563 571]{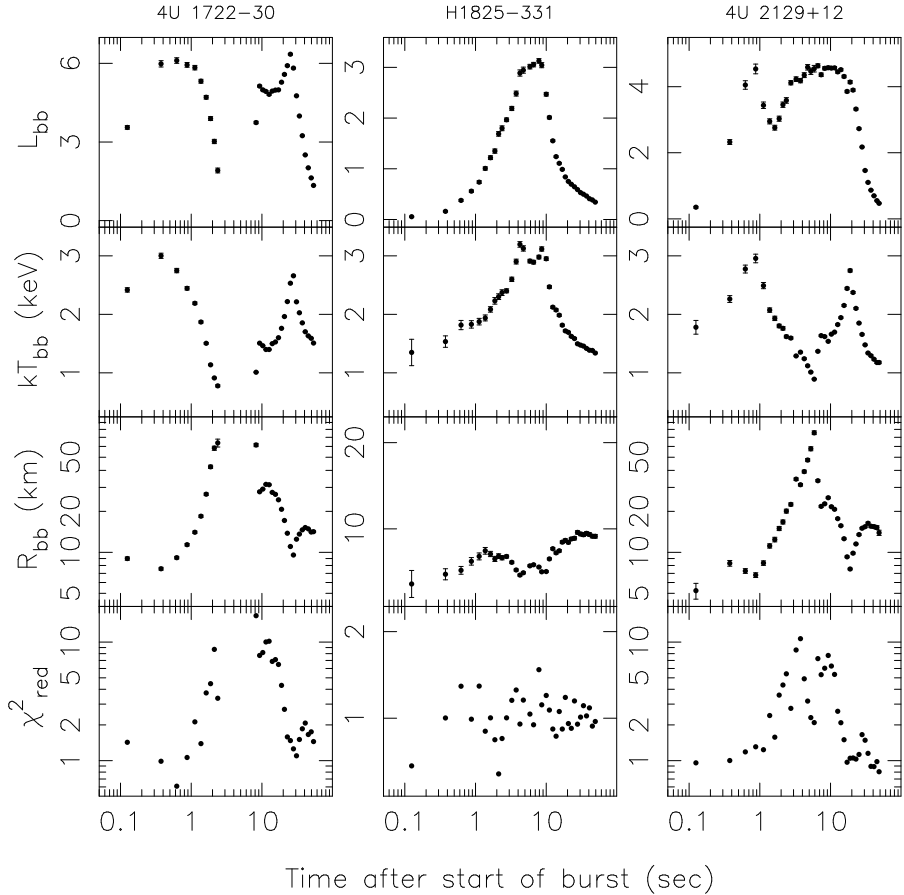}}
 \caption{Same as Fig.~\ref{plot_5RXTEbursts} but for three long X-ray bursts from three other
 globular cluster X-ray sources seen by the RXTE/PCA.
 Note that the R$_{\rm bb}$ and $\chi^2_{\rm red}$ axes for 4U\,1722$-$30 and 4U\,2129+12 
 have a logarithmic scale.
 T=0\,s corresponds to UT 1996 November 8 07:00:29, 1998 November 27 05:45:11 and 2000 September 22 13:47:37,
 for 4U\,1722$-$30, H1825$-$331 and 4U\,2129+12, respectively.
 \label{plot_pars_terzan2_ngc7078_ngc6652}
}
\end{center}
\end{figure*}

From the WFC data time resolved dead-time corrected spectra were determined for each X-ray burst. 
Different integration times were used, 
so that each spectrum had a similar signal to noise. This was 10 in most cases.
For a few X-ray bursts from 4U\,1722$-$30 and the only X-ray burst from 4U\,2129+12, 
however, this resulted in only a few time resolved spectra
in which case we could not determine whether it exhibited photospheric radius expansion or not. For these
we lowered the signal to noise down to 5. 
The X-ray bursts from 4U\,1820$-$30 typically last less than 25\,s with rather short photospheric radius 
expansion phases (see e.g.\ Fig.~2 and Fig.~5); for the X-ray burst spectra from this source, therefore, 
we also used a signal to noise of 5.

We performed
our spectral analysis (black-body radiation subjected to interstellar absorption) 
in the 3--20\,keV energy band for the RXTE/PCA and 2--20\,keV energy band for the BeppoSAX/WFC.
For all the X-ray bursts seen with the RXTE/PCA and BeppoSAX/WFC, we subtracted the 
pre-burst persistent emission from the X-ray burst spectra
(see e.g.\ Kuulkers et al.\ 2002, and references therein). The interstellar absorption, $N_{\rm H}$, 
was fixed to the values given in Table~\ref{GC}. 
These values are mostly from the best fitting models (disk black-body plus a Comptonized component 
[see Titarchuk 1994]) to wide-band X-ray spectral data from the BeppoSAX/NFI 
of the persistent emission by Sidoli et al.\ (2001) and Parmar et al.\ (2001). 
However, MXB\,1730$-$335 and XB\,1733$-$30 were too faint in their observations used, 
whereas no BeppoSAX/NFI observations have been done for XB\,1745$-$25.
For MXB\,1730$-$335 we, therefore, used the value of $N_{\rm H}$ from BeppoSAX/NFI results by Masetti et al.\ (2000), when the source was on, whereas for
XB\,1733$-$30 we used the value of $N_{\rm H}$ from power-law fit results to EXOSAT/ME spectra reported 
by Parmar et al.\ (1989);
the latter value of N$_{\rm H}$ is consistent 
with that reported from ROSAT/PSPC observations by Verbunt et al.\ (1995): 1.8$\times$10$^{22}$\,cm$^{-2}$.
For XB\,1745$-$25 we used the persistent source spectrum in outburst as observed by the BeppoSAX/WFCs to
constrain $N_{\rm H}$, see Appendix~\ref{ter5}; this value is also more or less consistent with that
derived from ROSAT/PSPC observations by Verbunt et al.\ (1995): 2$\times$10$^{22}$\,cm$^{-2}$.
Finally, the uncertainties quoted from the X-ray spectral fits are 1\,$\sigma$ errors, using 
$\Delta\chi^2=1$. 

\begin{figure*}
\begin{center}
 \resizebox{14.5cm}{!}{\includegraphics[angle=-90, clip, bb=31 37 282 769]{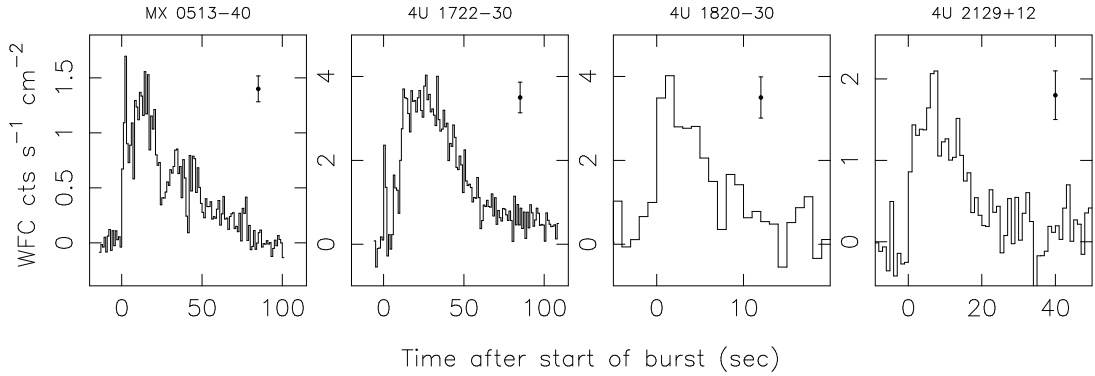}}
 \caption{BeppoSAX/WFC light curves (2--28\,keV) for four X-ray bursts from four globular cluster X-ray sources.
 The time resolution is 1\,s. T=0\,s correspond to UT 2000 September 11 20:01:54, 2000 March 6 23:59:51,
 1998 September 2 06:47:25, and 1997 May 9 09:00:18, for MX\,0513$-$40, 4U\,1722$-$30, 
 4U\,1820$-$30 and 4U\,2129+12, respectively.
 A representative error bar is given in the upper right of each panel. 
 The X-ray bursts seen from MX\,0513$-$40, 4U\,1722$-$30 and 4U\,1820$-$30 are radius expansion bursts, see Fig.~\ref{plot_pars_WFC}.
\label{plot_WFC_lc}
}
\end{center}
\end{figure*}

\begin{figure*}
\begin{center}
 \resizebox{14.5cm}{!}{\includegraphics[angle=-90, clip, bb=31 37 563 629]{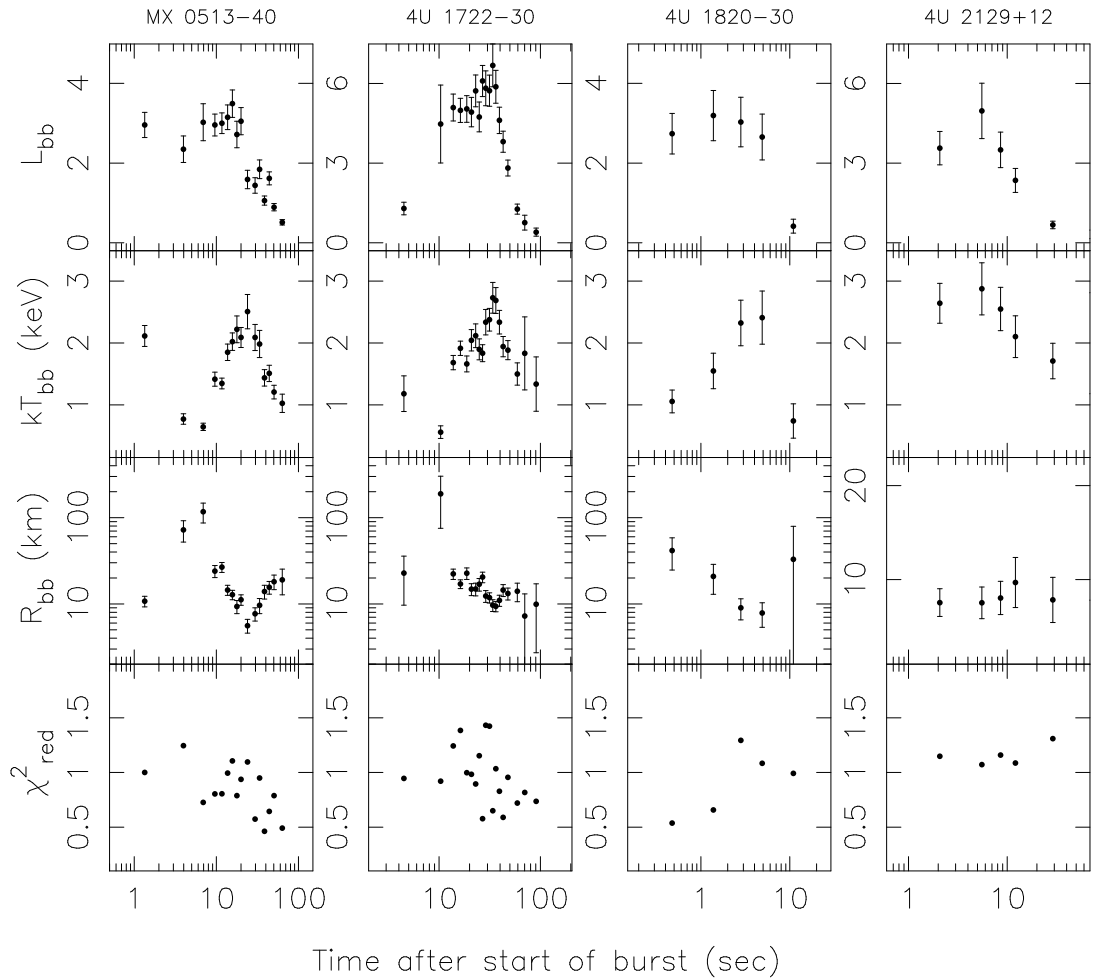}}
 \caption{Same as Fig.~\ref{plot_5RXTEbursts} but for four X-ray bursts from four globular cluster X-ray sources seen by the
 BeppoSAX/WFCs. For the light curves and the time of T=0\,s we refer to Fig.~4. 
 The values are also plotted on a logarithmic time scale, but have not been logarithmically rebinned in time.
 Note that the R$_{\rm bb}$ axis for MX\,0513$-$40, 4U\,1722$-$30 and 4U\,1820$-$30
 has a logarithmic scale.
 Note also that the first two points for 4U\,1722$-$30 are derived using spectra
 with a signal to noise of 5, in order to cover the beginning of the X-ray burst in
 more detail; the spectra for 4U\,2129+12 had a signal to noise of 6.
 \label{plot_pars_WFC}
}
\end{center}
\end{figure*}

\subsection{X-ray spectral analysis}

As noted in Sect.~1.1 the X-ray burst emission may not be entirely Planckian. Also,
a different spectral model has an impact on the inferred bolometric luminosity.
When determining the observed bolometric X-ray flux from our X-ray spectral fits
we do {\it not} take this into account, even when the X-ray spectral fits are satisfactory. 
In principle an assessment of the
observed bolometric X-ray flux using various spectral models is warranted
(see e.g.\ Damen et al.\ 1990). However, since
the `true' X-ray burst emission is still rather uncertain,
such an assessment will be somewhat elusive, and outside the scope of this paper.
We, therefore, assume that the X-ray spectra at maximum flux are Planckian and
that the unabsorbed bolometric X-ray flux may be determined using
\begin{equation}
F = \sigma T ( R / d )^2,
\end{equation}
where $T$ is the black-body temperature (or `color' temperature), $R$ the
neutron star photospheric radius, and $d$ the distance to the source.

\subsection{Literature values}

In Appendix~\ref{account} we give an account, focused on the bolometric X-ray burst peak fluxes,
of all the X-ray bursts seen from globular cluster X-ray sources presented
in the literature. These fluxes are also tabulated in Table~\ref{allbursts}.
The fluxes per source are ordered per instrument.
If an instrument did not see a radius expansion burst we give the maximum bolometric flux reached
during a single X-ray burst, or a range in bolometric peak fluxes if there is more than one 
X-ray burst observed with an instrument.
In the latter case sometimes also a value is quoted for individual (set of) X-ray bursts, i.e.\ either the maximum 
observed peak flux or the average bolometric peak flux of a subset.
We note that errors are not always given in the literature. 

\subsection{Cross calibration of different instruments}

It has been noted before that the spectral analysis of RXTE/PCA data of the Crab (pulsar plus nebula)
results in observed
fluxes which are about 20\%\ higher than those derived from Crab data taken with other instruments 
(see e.g.\ Tomsick et al.\ 1999; Wilms et al.\ 1999; Kuiper et al.\ 1999; Barret et al.\ 2000).
Although the calibration of the RXTE/PCA instrument has been improved, the absolute calibration
has still not been optimized (see e.g.\ Jahoda 2000). We, therefore, analysed data sets
of the Crab, in order to quantify the effect on our analysis. 
We created spectra using data recored by the {\sc Standard~2} mode, 
with either all PCUs, PCUs 0,1,2,3, or PCUs 0,2,3 summed. 
The analysis was done in the same way as described above for our globular cluster X-ray sources.
Background was subtracted from the Crab spectra, which was estimated using the 2002, February 26
background models. 
For the spectral model of the Crab (pulsar plus nebula) we used a power law subjected to
interstellar absorption:
\begin{equation}
N_{ph} = A \, E^{-\Gamma}e^{-N_{\rm H}\sigma(E)}, 
\end{equation}
where $N_{ph}$ is the flux in photons s$^{-1}$\,cm$^{-2}$\,keV$^{-1}$, $A$ the normalization in
photons s$^{-1}$\,cm$^{-2}$\,keV$^{-1}$ at 1\,keV, $E$ the photon energy, $\Gamma$ the power-law photon
index, $N_{\rm H}$ the interstellar hydrogen column density, and $\sigma$ the photo-electric absorption cross
section from Morrison \&\ McCammon (1983). 
$N_{\rm H}$ was fixed to 3$\times$10$^{21}$\,cm$^{-1}$.
Our spectral fit results to Crab data obtained on 2000, November 21 
are shown in Table~\ref{Crab}, together with the `canonical' Crab X-ray spectrum,
and the results from other X-ray instruments. For the instruments onboard SAS-3, Hakucho and
Granat (with which X-ray bursts were seen and which were used in our investigation) 
we do not have X-ray spectral information regarding the Crab.
The instruments onboard SAS-3 and Hakucho provided count rates in only a few energy channels.
In those cases it seems likely that they were calibrated by using the `canonical' Crab spectrum.

It is clear from Table~\ref{Crab} that the power-law normalization, $A$
(as well as the observed unabsorbed 3--20\,keV flux, $F_X$), derived for the RXTE/PCA
differs very significantly from those derived by other instruments. 
In order to account for this discrepancy we introduce a scale factor when
modeling the observed count rate spectra
(note that this results in a correction to the black-body flux and black-body radius, which we 
use in this paper).
These scale factors were determined by comparing the absorbed Crab
flux observed with the RXTE/PCA with the absorbed Crab flux estimated from the `canonical' Crab spectrum,
in the 3--20\,keV band (i.e., the energy range used in this paper). 
This is, of course, a crude correction, since the Crab spectral shape (power law)
differs from that of X-ray burst spectra (black body).
The scale factors\footnote{These scale factors have not been taken into account 
when quoting previous RXTE/PCA results in Appendix~\ref{account}.} are 
1.242, 1.211 and 1.227, respectively,
when all PCUs are used, only PCUs 0,1,2,3, or only PCUs 0,2,3.
After taking into account the correction 
factors in our RXTE/PCA spectral analyses, 
the final bolometric X-ray burst peak fluxes are more comparable to
those obtained with other instruments (see Table~\ref{allbursts}).

\section{Results}

\subsection{Maximum bolometric X-ray burst peak fluxes}
\label{candle}

The spectral fit results for the (brightest) X-ray bursts of 8 globular cluster sources seen by the RXTE/PCA are 
shown in Figs.~\ref{plot_5RXTEbursts} and \ref{plot_pars_terzan2_ngc7078_ngc6652}, for the short and long X-ray bursts, respectively. 
In Figs.~\ref{plot_WFC_lc} and \ref{plot_pars_WFC} we show the light curves and spectral fit results for X-ray bursts from four globular cluster sources seen by 
the BeppoSAX/WFCs. In Appendix~\ref{account} we describe these results in more detail.

During all strong radius expansion bursts seen with the RXTE/PCA 
(4U\,1722$-$30, 4U\,1820$-$30 and 4U\,2129+12) the spectral fits were rather poor during the photospheric radius expansion phase.
Our goodness of fit, expressed in $\chi^2_{\rm red}$ reached values of up to 25 for
13 degrees of freedom (dof) for 4U\,1722$-$30 (Fig.~\ref{plot_pars_terzan2_ngc7078_ngc6652}), 
up to 14 for 20 dof for 4U\,2129+12 (Fig.~\ref{plot_pars_terzan2_ngc7078_ngc6652}), 
and up to 7 for 23 dof for 4U\,1820$-$30 (Fig.~\ref{plot_5RXTEbursts}). 
Also in the case of XB\,1745$-$25 the spectral fits are somewhat poor during the photospheric radius expansion phase
($\chi^2_{\rm red}$ of up to 2 for 18 dof; see Fig.~\ref{plot_5RXTEbursts}).
The high values of $\chi^2_{\rm red}$ were not reported by Molkov et al.\ (2000) and Smale (2001) 
in the former two sources, 
whereas it was reported for 4U\,1820$-$30 (Franco \&\ Strohmayer 1999). 
The values of $\chi^2_{\rm red}$ reduce somewhat if one uses smaller
energy range in the spectral fits (as was done by Smale [2001]: 2.5--15\,keV).
In this respect it is interesting to note that in the early part of the X-ray burst 
(which contains the photospheric radius expansion/contraction phase) from A1850$-$08
black-body models did not provide acceptable fits, whereas they were acceptable during
the cooling stage (Hoffman et al.\ 1980).
In most of the cases the RXTE/PCA spectra exhibited residuals which were
similar to that reported during the photospheric radius expansion phases of the 
strong X-ray burst from 4U\,2129+12 seen by the Ginga/LAC (van Paradijs et al.\ 1990), 
the long X-ray bursts seen from GX\,17+2 (Kuulkers et al.\ 2002), and the superburst from 4U\,1820$-$30 
(Strohmayer \&\ Brown 2002).
Additional components are necessary to improve the fits, such as a line and an edge feature
(see Strohmayer \&\ Brown 2002). Moreover, the emission during the X-ray burst may be 
non-Planckian (see Sect.~1.1).
However, we are only interested in the maximum observed flux, which
is most of the time reached near the end of the photospheric radius expansion phase where the spectral fits are
good ($\chi^2_{\rm red}$$\sim$1).
We note that some of the BeppoSAX/WFC spectral fits 
reached rather low values of $\chi^2_{\rm red}$
(down to about 0.5 for 23 dof). We ascribe this to a possible overestimate of the systematic
uncertainties involved in the BeppoSAX/WFC analysis.

We recorded the peak bolometric black-body fluxes reached during the X-ray bursts in our sample, 
from the various spectral fits. The result can be found in Table~\ref{allbursts}, together
with the results from the literature. 
Note that one expects the maximum flux to be approximately constant when
radiating near the Eddington limit during the photospheric radius expansion stage.
As shown above, in most cases this is not observed, and the maximum flux is often
seen when the photospheric radius has shrunk back to its pre-burst value.
This is only partly due to the fact that at large photospheric radii
the temperature is low (typically $\lesssim$1\,keV, see also Sect.~1.1).
Most of the X-ray emission is then outside the observed energy range,
making the bolometric correction uncertain (see e.g.\
Lewin et al.\ 1984; van Paradijs et al.\ 1990).
Our observed peak values may therefore not represent the 
actual peak fluxes of the X-ray bursts, and some caution must be taken.
Also, because
of the different statistical quality of the data from different instruments, 
the accumulated spectra from which the peak fluxes are determined, may have different integration times.
The averaged peak fluxes derived from the spectra with longer time bins may, therefore, be 
underestimated, when the photospheric radius expansion phase lasts shorter than the timebin 
or if there is appreciable variability during the photospheric radius expansion phase.
However, we consider this effect to be small or even negligible for most of the X-ray bursts in our final sample. 

\subsection{Do X-ray bursts of one system reach the same peak flux?}

\begin{figure}
\begin{center}
 \resizebox{7cm}{!}{\includegraphics[angle=-90, clip, bb=62 91 571 399]{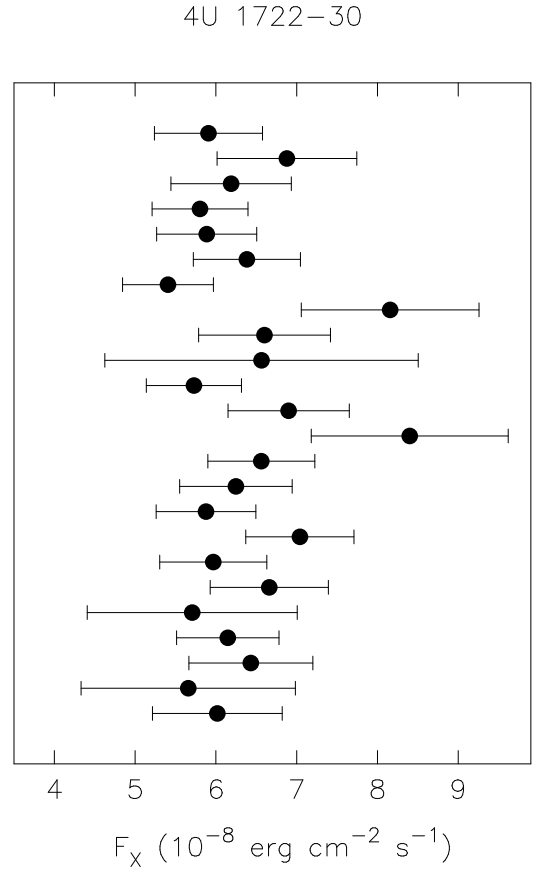}}
 \caption{Bolometric peak fluxes, F$_{\rm X}$, reached during the 24 radius expansion bursts from
 4U\,1722$-$30 as seen by the BeppoSAX/WFCs. 
 They are all consistent with one value, 6.23$\pm$0.15$\times$10$^{-8}$\,erg\,cm$^{-2}$\,s$^{-1}$, see 
 text.
\label{plot_ter2}
}
\end{center}
\end{figure}

The X-ray burst bolometric peak fluxes of the 24 radius expansion bursts from 4U\,1722$-$30 seen by the BeppoSAX/WFC 
are shown in Fig.~\ref{plot_ter2}. 
They are consistent with one value ($\chi^2_{\rm red}$=0.64 for 23 degrees of freedom):
the weighted average bolometric peak flux is
6.23$\pm$0.15$\times$10$^{-8}$\,erg\,cm$^{-2}$\,s$^{-1}$. 
A similar analysis of 21 X-ray bursts seen by the BeppoSAX/WFCs from 4U\,1820$-$30 showed
15 radius expansion bursts (see e.g.\ Fig.~5). 
Their bolometric peak fluxes are consistent with one value of 
5.9$\pm$0.3$\times$10$^{-8}$\,erg\,cm$^{-2}$\,s$^{-1}$
($\chi^2_{\rm red}$=0.82 for 14 degrees of freedom).
This shows that the peak fluxes of the X-ray bursts from 4U\,1722$-$30 and 4U\,1820$-$30
are each compatible with a single value.

If one compares the results from different reports of the same X-ray burst(s), 
one sees that the reported bolometric peak fluxes are
somewhat different (4U\,1722$-$30, 4U\,1820$-$30, A1850$-$08 and 4U\,2129+12). 
When comparing bolometric 
peak fluxes in radius expansion bursts 
from different instruments, the differences are also apparent
(4U\,1722$-$30, 4U\,1820$-$30 and 4U\,2129+12). 
Since for multiple radius expansion bursts
seen with the same instrument the bolometric peak fluxes were found to be more or less
consistent with each other,
we attribute it to 
the differences in the response matrices used (usually the calibration of an instrument improves with time),
in the case of the same instrument, and to the differences in the absolute calibration, when
comparing different instruments.
This was shown to be the case for the RXTE/PCA where the calibration information
is readily available (see Sect.~3.5). 

\subsection{Do X-ray bursts from different systems reach the same peak flux?}

\begin{figure*}
\begin{center}
 \resizebox{12cm}{!}{\includegraphics[angle=-90, clip, bb=80 52 565 550]{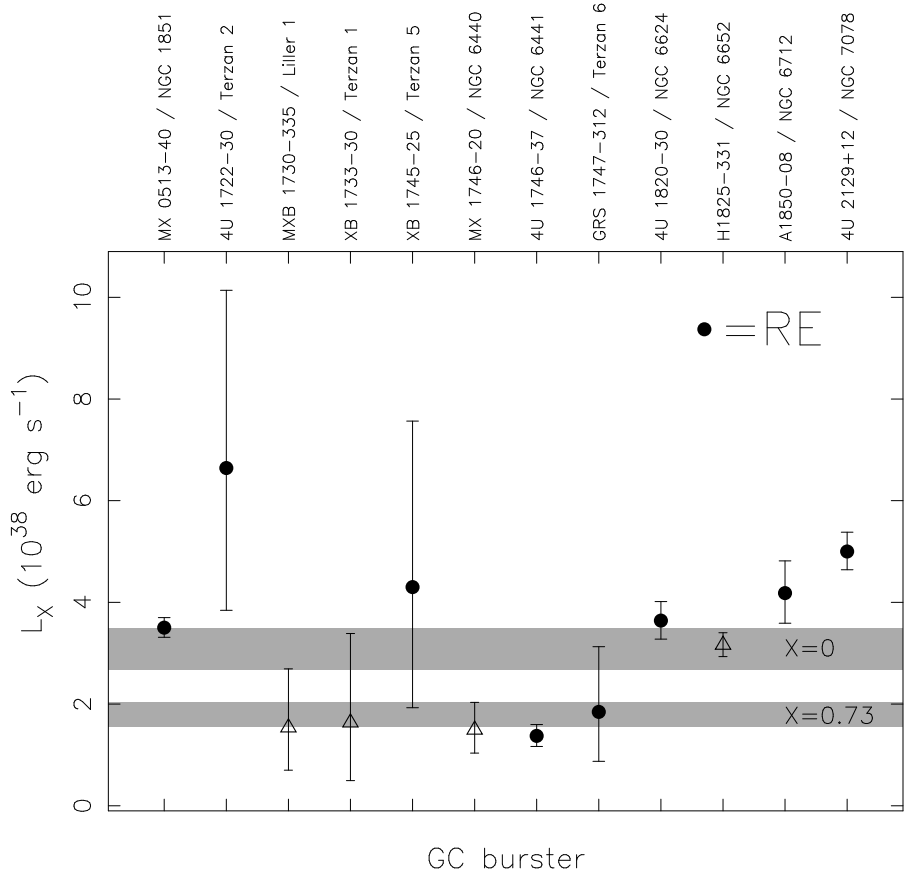}}
 \caption{
 Bolometric peak luminosities reached during the brightest X-ray bursts seen in the twelve globular
 cluster X-ray sources, using the distances and bolometric peak fluxes as quoted in Tables~1 and 2. 
 Photospheric radius expansion (RE) X-ray bursts have been denoted with a filled circle; the
 others with an open triangle. With grey bands we give 
 the expected range in the Eddington limit for matter with cosmic composition 
 (X=0.73) and hydrogen-poor matter (X=0), see Sect.~5.
\label{plot_lx}
 }
\end{center}
\end{figure*}

We derived the weighted average X-ray burst bolometric peak 
flux per source using the peak fluxes from individual or a subset of radius expansion bursts. 
The corresponding uncertainties are either the uncertainty in a single value or in the mean of
two bolometric flux values from the same instruments, or the spread in the bolometric 
flux values from various instruments.
This partly alleviates the calibration problems discussed above.
The bolometric peak flux values used are indicated with a $\surd$ in Table~\ref{allbursts}.
For those sources with multiple X-ray bursts for which no radius expansion burst has been observed
(MXB\,1730$-$335, XB\,1733$-$30 and H1825$-$331),
we used the highest observed bolometric peak fluxes;
they are also indicated with a $\surd$ in Table~\ref{allbursts}.

The resulting bolometric peak fluxes are given in Table~\ref{GC}. Using the distances as given in the 
Table~\ref{Harris}, we determine the corresponding bolometric peak luminosities (assuming isotropic emission). 
We propagate the errors in the distances and bolometric fluxes, assuming they are independent.
In Fig.~7 we plot these peak luminosities.
Two third of our sample show radius expansion bursts which reach a 
bolometric peak luminosity roughly 
between 1.5 and 7$\times$10$^{38}$\,erg\,s$^{-1}$, with a weighted mean of 
3.01$\pm$0.12$\times$10$^{38}$\,erg\,s$^{-1}$
($\chi^2_{\rm red}$=14.6 for 7 dof; assuming symmetric errors).
Excluding 4U\,1746$-$37 (whose radius expansion bursts reach a significantly lower peak value
than most of the radius expansion bursts from the other sources; see also Sect.~5)
we obtain a weighted mean value of 3.79$\pm$0.15$\times$10$^{38}$\,erg\,s$^{-1}$, with
$\chi^2_{\rm red}$=2.9 for 6 dof (assuming symmetric errors). 
Thus, when taking the peak luminosities at face value, 
the hypothesis that the peak luminosities
of X-ray bursts with photospheric radius expansion is the same for all sources
must be rejected. Indeed, Fig.~7 shows that the peak luminosities
of mainly MX\,0513$-$40 and 4U\,2129+12
are significantly different (excluding 4U\,1746$-$37). 
Note that the peak luminosity reached during the X-ray burst of H1825$-$331 is close
to the mean peak luminosity (Fig.~7); this confirms our suggestion from the spectral fit parameters
that it was on the verge of showing photospheric radius expansion (see Appendix~B.10).

\section{Discussion}
\label{discussion}

We find that the photospheric radius expansion type~I X-ray bursts from bright X-ray sources in globular clusters
reach an empirical bolometric peak luminosity 
(assuming the total X-ray burst emission is entirely 
due to black-body radiation and the recorded maximum luminosity is the actual peak luminosity) 
of roughly 3.8$\times$10$^{38}$\,erg\,s$^{-1}$
(excluding 4U\,1746$-$37, see also below). Because significant deviations from this value occur
(e.g., 4U\,2129+12) this only approximately 
confirms an earlier suggestion by Basinska et al.\ (1984) that radius expansion bursts reach 
a `true' critical luminosity. 
Our value for the average bolometric peak luminosity is very close to the `adopted 
standard candle' value of 3.7$\times$10$^{38}$\,erg\,s$^{-1}$ 
by Verbunt et al.\ (1984) and comparable to the mean bolometric peak luminosity of
3.0$\pm$0.5$\times$10$^{38}$\,erg\,s$^{-1}$ by Lewin et al.\ (1993) 
of X-ray bursts with photospheric radius expansion from only four globular
cluster bursters (i.e., 4U\,1722$-$30, 4U\,1820$-$30, A1850$-$08 and 4U\,2129+12).

To establish whether radius expansion bursts can be used as standard candles, one first has
to verify whether the X-ray burst fluxes during every radius expansion burst in an individual source
reach a single maximum value.
Observations prior to RXTE and BeppoSAX revealed many (i.e., typically more than ten) 
radius expansion bursts in one globular cluster source, 4U\,1820$-$30 (Vacca et al.\ 1986; 
Haberl et al.\ 1987; Damen et al.\ 1990; see Appendix~\ref{account}), and two sources
outside globular cluster sources,
MXB\,1636$-$53 (Inoue et al.\ 1984; Damen et al.\ 1990) and MXB\,1728$-$34 (Basinska et al.\ 1984).
The bolometric peak fluxes were found to be the same within the errors in all three cases.

The BeppoSAX/WFCs observed many radius expansion bursts from the globular cluster sources 4U\,1820$-$30
and 4U\,1722$-$30. We find that the bolometric peak fluxes are also the same within the errors in both
cases. 
A similar conclusion was reached from eight radius expansion bursts seen by the BeppoSAX/WFCs 
from one source outside a globular cluster, XB\,1812$-$12 (Cocchi et al.\ 2000a).
The more sensitive RXTE/PCA observations of especially MXB\,1728$-$34 
(e.g.\ van Straaten et al.\ 2001; Franco 2001)
and MXB\,1636$-$53 show, however, that the bolometric peak fluxes are not exactly constant, with 
standard deviations of $\simeq$9.5\%\ and $\simeq$7\%, respectively 
(Galloway et al.\ 2002a; 2002b). Similar studies in other sources 
but with less radius expansion bursts (between four to eight per source) indicate standard deviations between 
7\%\ and 15\%\ (Muno et al.\ 2000; Kuulkers et al.\ 2002; Galloway et al.\ 2002a; 2002b).
In the case of MXB\,1728$-$34 it was found that the peak fluxes vary systematically on a monthly timescale; taking this
secular variation into account the residual scatter is only $\simeq$3\%\ (Galloway et al.\ 2002a; 2002b). 

For the globular cluster bursters MX\,0513$-$40, XB\,1745$-$25 and 4U\,1746$-$37 only two radius expansion bursts 
were seen from one instrument. In all three cases similar peak fluxes within the errors were reached for both X-ray bursts 
(see Appendix~\ref{account}). 

We note that two sources outside globular clusters, MXB\,1659$-$29 (Wijnands et al.\ 2002) 
and EXO\,0748$-$676 (Gottwald et al.\ 1986), showed very different peak fluxes 
during different radius expansion bursts.
They both are viewed at a high inclination
(e.g.\ Cominsky \&\ Wood 1984; Parmar et al.\ 1986), 
so that, although the true critical luminosity probably is reached, 
the X-ray burst emission may be partly obscured by the accretion disk and/or donor star
(i.e.\ the X-ray burst emission is anisotropic). This may lead to different {\it observed} maximum peak luminosities
at different times.
Note that this may also affect the peak luminosities during the radius expansion bursts
from the globular cluster bursters 4U\,1746$-$37 and GRS\,1747$-$312, since they are also
relatively high inclination systems (e.g.\ Sansom et al.\ 1993; in~'t~Zand et al.\ 2000; see also
Sztajno et al.\ 1987, who concluded from other grounds that the X-ray burst emission 
from 4U\,1746$-$37 is probably anisotropic).

Excluding the high-inclination systems, we conclude 
that, since multiple radius expansion bursts in a single source reach the same 
maximum luminosity to within $\sim$10\%\ and since 
the maximum luminosity is the same within $\sim$15\%\ for different globular cluster sources, the
`true' critical luminosity can only be regarded as an approximate empirical
standard candle. 

In Fig.~7 we show in grey bands the range in the Eddington luminosity limit for matter with 
cosmic composition 
(X=0.73; lower band) and hydrogen-poor matter (X=0; upper band) for a (canonical)
1.4\,M$_{\odot}$ neutron star with a radius of 10\,km. 
The upper and lower boundaries of each band is determined by
the Eddington limit observed by an observer far away from the neutron star with very strong photospheric radius expansion
and very weak photospheric radius expansion, respectively (van Paradijs 1979; Goldman 1979; see Lewin et al.\ 1993). 
Note that an increase in the neutron star mass shifts the band slightly upwards 
as well as substantially broadens it.
A higher than canonical neutron star mass seems to be required in some models
explaining the kHz QPO in neutron star low-mass X-ray binaries (see e.g.\ Stella et al.\ 1999;
Stella 2001; Lamb \&\ Miller 2001), and is derived dynamically for the 
low-mass X-ray binary and 
X-ray burster Cyg\,X-2 (Orosz \&\ Kuulkers 1999).
Our critical luminosity is close to the Eddington luminosity
limit for hydrogen-poor matter. Therefore, 
our critical luminosity is most likely this Eddington limit (see also e.g.\ van Paradijs 1978).

Previously, `super-Eddington luminosities' were reported for several sources in our sample
(e.g.\ Grindlay et al.\ 1980; van Paradijs 1981; Inoue et al.\ 1981; see also Verbunt et al.\ 1984)
and it was referred to as the `super Eddington limit problem' (e.g.\ Lewin 1985).
Winds or outflows from the neutron-star surface set-up by the high
radiation pressure were introduced to resolve (part of) this problem
(e.g.\ Melia \&\ Joss 1985; Stollman \&\ van Paradijs 1985).
Our findings do not support super-Eddington luminosities, except possibly for 4U\,2129+12
(note that 4U\,2129+12 {\it is} consistent with the Eddington limit for hydrogen-poor matter if
we assume e.g.\ a somewhat higher neutron star mass, see above).
We attribute this mainly
to a better knowledge of the distance and their uncertainties.

Five bursters (including GRS\,1747$-$312)
in our sample show bolometric peak X-ray burst luminosities near 1.5$\times$10$^{38}$\,erg\,s$^{-1}$, which
is comparable to the Eddington limit of matter with solar composition. 
Of these, only the sources 4U\,1746$-$37 and GRS\,1747$-$312 showed X-ray bursts with evidence 
for photospheric radius expansion; the maximum peak X-ray burst luminosity for the other sources could have been higher if 
radius expansion bursts were seen. 

If there is a bimodal distribution of the bolometric X-ray burst peak fluxes it 
is consistent with the thought that the donors in systems with an ultra-short orbital period 
($\lesssim$1\,hr: MX\,0513$-$40, H1825$-$331, 4U\,1820$-$30, A1850$-$08; see Table~\ref{GC})
do not provide hydrogen, whereas the donors in systems with a longer orbital period 
(4U\,1746$-$37, GRS\,1747$-$312; see Table~\ref{GC}\footnote{The
17.1\,hr orbital period found previously for 4U\,2129+12
(Ilovaisky et al.\ 1993) is {\it not} attributed to the
burster but to AC~211 (White \&\ Angelini 2001).}) do provide
hydrogen-rich material (e.g.\ Nelson et al.\ 1986; Verbunt 1987; see also
Juett et al.\ 2001). The short duration of the X-ray bursts from XB\,1745$-$25 and 4U\,1820$-$30 ($\lesssim$25\,s) 
is consistent with the fuel being only helium (Bildsten 1995). However, the X-ray bursts with longer duration 
($\sim$mins) from the other systems can not be accomodated this way (e.g.\ Fujimoto et al.\ 1981; 
Bildsten 1998, and references therein), unless the mass accretion rate onto the neutron star
is rather low (see Bildsten 1995). On the other hand, we note that the Eddington limit of hydrogen poor matter
may still be reached during long X-ray bursts if a hydrogen rich envelope is ejected during the photospheric radius
expansion stage of a helium flash (Sugimoto et al.\ 1984).

\begin{acknowledgements}
We thank Serge Molkov and Alan Smale for sending their 
X-ray burst spectral parameter fit results of 4U\,1722$-$30 and 4U\,2129+12, respectively, in numerical form,
Giancarlo Cusumano for providing the X-ray spectral fit results on Crab observations with the 
BeppoSAX/MECS, Duncan Galloway for sharing some of the RXTE/PCA results on radius expansion bursts
from non-globular cluster bursters, and Marten van Kerkwijk for providing comments on an earlier draft. 
This research has made use of the SIMBAD database, operated at CDS, Strasbourg, France.
The BeppoSAX satellite is a joint Italian and Dutch programme. 
\end{acknowledgements}

\appendix

\section{Interstellar absorption towards XB\,1745$-$25}
\label{ter5}

A value of $N_{\rm H}$ was reported for XB\,1745$-$25 in Terzan 5 on the basis of
ROSAT/PSPC observations (Verbunt et al.\ 1995): 2$\times$10$^{22}$\,cm$^{-2}$. 
The BeppoSAX/WFCs observed XB\,1745$-$25 through the various stages of its
2000 outburst for a net exposure of 761\,ksec. The source was quite bright (about 0.5\,Crab), allowing
a good measure of $N_{\rm H}$, since the absorption is already visible at
$\simeq$4\,keV. We accumulated a dead-time corrected spectrum using data 
from WFC unit 2 obtained between 
UT 2000 August 8 20:44:38 to August 9 20:53:48, giving an effective exposure time of 31.4\,ksec. 

We fitted the WFC spectrum with similar models as used by Sidoli et al.\ (2001), as well
as a black body plus a power-law. The fit results are displayed in Table~\ref{tabter5}.
The disk black body plus Comptonization model provided the best fits, similar to the results 
by Sidoli et al.\ (2001). We therefore use the value of $N_{\rm H}$ from this model
for our analysis, see Table~\ref{GC}, which is somewhat higher than that found by Verbunt et al.\ (1995). 
The mean persistent flux was $\simeq$1.16$\times$10$^{-8}$\,erg\,cm$^{-2}$\,s
(2--20\,keV).

\begin{table*}
\caption{X-ray spectral fits results to the persistent emission of XB\,1745$-$25 during outburst observed
with the BeppoSAX/WFCs. Three models were used; the first two (disk black body [{\sc diskbb} in {\sc XSPEC}] plus
a Comptonization component [{\sc comptt}, see Titarchuk 1994] and black body [{\sc bbodyrad}] plus a Comptonization component) 
were similar to those studied by Sidoli et al.\ (2001). The third (black body plus a power law) is used for 
comparison. 
The {\sc diskbb} model has as free parameters: $kT_{\rm in}$, the temperature at the inner disk radius,
and $r_{\rm in}/d_{10}(\cos{i})^{0.5}$, where $r_{\rm in}$ is the inner disk radius, $d_{10}$ the distance to the source
in units of 10\,kpc, and $i$ the inclination angle of the disk.
The {\sc comptt} model has as free parameters: $kT_e$, the temperature of the Comptonizing electrons, $\tau_p$,
the plasma optical depth with respect to electron scattering, $kT_0$, the input temperature of the soft photon
(Wien) distribution, and a normalization constant. 
The {\sc bbodyrad} model has as free parameters: $kT_{\rm bb}$, the black-body temperature, and
$r_{\rm bb}$, the black-body radius for a source at 10\,kpc. 
The {\sc powerlaw} model has as free parameters:
$\Gamma$, the power-law photon index, and the power-law normalization in
photons\,keV$^{-1}$\,cm$^{-2}$\,s$^{-1}$ at 1\,keV.
}
\begin{tabular}{ccccccccc}
\hline
\multicolumn{9}{c}{~} \\
 model   & $N_{\rm H}$            & $kT_{\rm in}$ & $r_{\rm in}/d_{10}(\cos{i})^{0.5}$ & $kT_0$ & $kT_e$ & $\tau_p$ & norm & $\chi^2_{\rm red}$/dof \\ 
              & (10$^{22}$\,cm$^{-2}$) & (keV)         & (km)                             & (keV)  & (keV)  &          & & \\
{\sc diskbb+comptt} & 3.8$^{+0.9}_{-0.7}$ & 0.60$^{+0.07}_{-0.04}$ & 72$^{+57}_{-21}$ & 1.28$^{+0.06}_{-0.09}$ & 3.7$^{+1.0}_{-0.5}$ & 3.6$^{+0.8}_{-0.9}$ & 0.57$^{+0.12}_{-0.14}$ & 1.15/21 \\
\multicolumn{9}{c}{~} \\
\hline
\multicolumn{9}{c}{~} \\
model    & $N_{\rm H}$            & $kT_{\rm bb}$ & $r_{\rm bb}$ & $kT_0$ & $kT_e$ & $\tau_p$ & norm & $\chi^2_{\rm red}$/dof \\ 
         & (10$^{22}$\,cm$^{-2}$) & (keV)         & (km)             & (keV)  & (keV)  &          & & \\
{\sc bbodyrad+comptt}    & 2.3$^{+0.6}_{-0.8}$ & 1.69$^{+0.07}_{-0.08}$ & 8.3$^{+0.8}_{-0.5}$ & 0.35$^{+0.10}_{-0.31}$ & 4.3$^{+2.3}_{-0.8}$ & 4.2$^{+1.3}_{-1.4}$ & 0.7$^{+0.7}_{-0.2}$ & 1.38/21 \\
\multicolumn{9}{c}{~} \\
\hline
\multicolumn{9}{c}{~} \\
 model   & $N_{\rm H}$            & $kT_{\rm bb}$ & $r_{\rm bb}$ & $\Gamma$      & norm & $\chi^2_{\rm red}$/dof & & \\ 
         & (10$^{22}$\,cm$^{-2}$) & (keV)         & (km)         & &  & & & \\
{\sc bbodyrad+powerlaw}   & 3.4$\pm$0.3 & 1.79$\pm$0.02 & 7.9$\pm$0.2      & 2.63$^{+0.06}_{-0.07}$ & 6.1$^{+0.9}_{-0.8}$ & & & 1.40/23 \\
\multicolumn{9}{c}{~} \\
\hline
\end{tabular}
\label{tabter5}
\end{table*}

\section{An account of all X-ray bursts seen from globular cluster bursters}
\label{account}

In this appendix we give an account of all X-ray bursts reported previously, as well
as those analysed by us. 
It is organized as follows: the first paragraph of each subsection describes the results 
from the literature, whereas in the second paragraph we describe our RXTE/PCA and/or BeppoSAX/WFC
results (if present). 
At the end we summarize in Table~\ref{allbursts} the bolometric 
peak fluxes for all the X-ray bursts; 
the peak fluxes used in our investigation are indicated with a $\surd$.
The RXTE/PCA peak fluxes quoted from the literature are not corrected for the absolute RXTE/PCA calibration 
(Sect.~3.5). The correction has been applied to our derived peak fluxes from the RXTE/PCA data; 
in Table~\ref{allbursts} we give between brackets also the uncorrected peak fluxes.

\subsection{MX\,0513$-$40/NGC\,1851}

The first X-ray burst from MX\,0513$-$40 was discovered by Uhuru (Forman \&\ Jones 1976). It had a peak
flux of about 0.5$\times$10$^{-8}$\,erg\,cm$^{-2}$\,s$^{-1}$, but the start of the X-ray burst was not seen.
Five X-ray bursts were seen by SAS-3/HTC (see Li \&\ Clark 1977), four of which had
bolometric peak fluxes ranging from 0.9$\pm$0.3 to 
1.7$\pm$0.3$\times$10$^{-8}$\,erg\,cm$^{-2}$\,s$^{-1}$ (Cominsky 1981). More recently, an X-ray burst was
seen with the Chandra/HRC-S (Homer et al.\ 2001). So far no radius expansion bursts had been
reported.

The RXTE/PCA observed one short ($\lesssim$25\,s) X-ray burst from MX\,0513$-$40. It did not exhibit
photospheric radius expansion. The BeppoSAX/WFCs observed eleven X-ray bursts from this source. The longest
one is shown in Fig.~\ref{plot_WFC_lc}. This X-ray burst showed a clear strong (factor $\sim$10) photospheric radius expansion phase;
its bolometric peak flux of about 2$\times$10$^{-8}$\,erg\,cm$^{-2}$\,s$^{-1}$ is reached near touch-down
(Fig.~\ref{plot_pars_WFC}). Another X-ray burst showed some evidence for (weak; factor $\sim$2) photospheric radius expansion.
This is the first time radius expansion bursts are reported from this source.

\subsection{4U\,1722$-$30/Terzan 2}

A long ($\gtrsim$10\,min) X-ray burst was observed from the direction of 4U\,1722$-$30 by OSO-8
(Swank et al.\ 1977). It reached a peak flux of about 6.2$\times$10$^{-8}$\,erg\,cm$^{-2}$\,s$^{-1}$.
During the first 20\,s of the event the black-body temperature was lowest, while the inferred black-body radius was 
100$^{+60}_{-20}$\,km (at 10\,kpc). During the remainder of the X-ray burst the inferred black-body radius was
15$\pm$4.5\,km (at 10\,kpc). This indicates that it was a (strong) radius expansion burst. 
Other strong (factor $\gtrsim$10) radius expansion bursts were seen by Granat/ART-P and RXTE/PCA (Molkov et al.\ 2000),
whereas a weak radius expansion burst (factor $\sim$2) was reported from the Einstein/MPC (Grindlay et al.\ 1980).
The RXTE/PCA event reached a bolometric peak flux of about 
7.29$\pm$0.06$\times$10$^{-8}$\,erg\,cm$^{-2}$\,s$^{-1}$, whereas
the event seen with the Einstein/MPC reached a bolometric peak flux of about 5.2$\times$10$^{-8}$\,erg\,cm$^{-2}$\,s$^{-1}$, 
both near touch down. 
Another X-ray burst was reported from BeppoSAX/NFI data; no clear photospheric radius expansion was observed.
The peak flux was $\simeq$5.0$\times$10$^{-8}$\,erg\,cm$^{-2}$\,s$^{-1}$ (Guainazzi et al.\ 1998).
The BeppoSAX/WFCs observed fourteen strong radius expansion bursts in the first three years of monitoring the
Galactic Center region (Cocchi et al.\ 2000b). Another ten were seen in later observations, see below.
The peak intensities of four of the fourteen X-ray bursts were consistent 
with each other, with a weighted mean of 1.01$\pm$0.03~Crab, which was extrapolated to a bolometric
flux of 3.2$\pm$0.1$\times$10$^{-8}$\,erg\,cm$^{-2}$\,s$^{-1}$. One of these X-ray bursts was 
analysed in more detail; from their spectral fits we infer that a bolometric peak flux of about 
3.5$\times$10$^{-8}$\,erg\,cm$^{-2}$\,s$^{-1}$ was reached (note that Cocchi et al.\ 2001 quote a
peak flux of 4.1$\pm$0.4$\times$10$^{-8}$\,erg\,cm$^{-2}$\,s$^{-1}$). 

Our fit parameter values for the X-ray burst from 4U\,1722$-$30 seen by RXTE/PCA (Fig.~3) are in agreement with those 
reported by Molkov et al.\ (2000). It is a very strong (factor $>$10) radius expansion burst; the maximum 
bolometric flux was reached  
near touch-down (Fig.~\ref{plot_pars_terzan2_ngc7078_ngc6652}). 
The BeppoSAX/WFCs observed a total of 24 X-ray bursts. Most of them have a `precursor' event in the lightcurves
(see e.g.\ Fig.~\ref{plot_WFC_lc}), indicating strong photospheric radius expansion (see Fig.~\ref{plot_pars_WFC}). 
All the X-ray bursts are radius expansion bursts. We found maximum bolometric peak fluxes between 5.4$\pm$0.6 and
8.4$\pm$1.2$\times$10$^{-8}$\,erg\,cm$^{-2}$\,s$^{-1}$, with a mean value
of 6.23$\pm$0.15$\times$10$^{-8}$\,erg\,cm$^{-2}$\,s$^{-1}$ (see also Sect.~\ref{candle}).
For the X-ray burst analysed in detail by
Cocchi et al.\ (2001; see above) 
we found a bolometric peak flux of 6.2$\pm$0.6$\times$10$^{-8}$\,erg\,cm$^{-2}$\,s$^{-1}$.
We attribute the difference with the lower flux reported by Cocchi et al.\ (2001) 
to an improved response matrix (in~'t~Zand 1999; see also in~'t~Zand et al.\ 2001a). Also, 
our values for the peak fluxes are comparable with those derived for other instruments. We therefore 
do not quote the earlier results obtained by the BeppoSAX/WFCs in Table~\ref{allbursts}.

\subsection{MXB\,1730$-$335/Liller 1}

MXB\,1730$-$335 (The Rapid Burster) is a source which receives a lot of attention. It is a transient source; 
in the first weeks of an outburst it shows a phase during which only
type~I X-ray bursts are seen. The rest of the outburst is dominated by type~II X-ray bursts, with occasionally
a type~I X-ray burst (Guerriero et al.\ 1999). Most authors focus on the phenomenology of the type~II
X-ray bursts. No radius expansion burst has so far been reported. The average type~I X-ray burst peak flux is about 
1.7$\times$10$^{-8}$\,erg\,cm$^{-2}$\,s$^{-1}$ (SAS-3; Marshall et al.\ 1979). 
For a more detailed account of the properties of MXB\,1730$-$335
we refer to Lewin et al.\ (1993), Guerriero et al.\ (1999), 
and Masetti et al.\ (2000), and references therein.

We selected various (mostly) bright X-ray bursts observed by RXTE/PCA from MXB\,1730$-$335, as reported by 
Guerriero et al.\ (1999) and seen in later outbursts. None of them showed photospheric radius expansion.
The brightest X-ray bursts show typical bolometric 
peak fluxes in the range $\simeq$1--2$\times$10$^{-8}$\,erg\,cm$^{-2}$\,s$^{-1}$, consistent with
values reported previously.
The fit results for one of the brightest X-ray bursts are shown in Fig.~\ref{plot_5RXTEbursts}.

\subsection{XB\,1733$-$30/Terzan 1}

XB\,1733$-$30 has been off for some time now (see e.g.\ Guainazzi et al.\ 1999). 
Only three short ($\lesssim$25\,s) X-ray bursts have been reported in the past. 
Two X-ray bursts were discovered by the Hakucho/FMC (Makishima et al.\ 1981). They reached an average maximum peak flux of 
7.4$\pm$1.0$\times$10$^{-8}$\,erg\,cm$^{-2}$\,s$^{-1}$ (Inoue et al.\ 1981). 
Another X-ray burst was observed by Granat/ART-P; 3--5\,s into the X-ray burst a bolometric peak flux of 
6.3$\pm$0.8$\times$10$^{-8}$\,erg\,cm$^{-2}$\,s$^{-1}$ was found (Pavlinsky et al.\ 2001).
No photospheric radius expansion phase was reported.

\subsection{XB\,1745$-$25/Terzan 5}

Fourteen X-ray bursts from the transient XB\,1745$-$25 were discovered by Hakucho (Makishima et al.\ 1981).
The maximum peak flux was 6.1$\pm$0.7$\times$10$^{-8}$\,erg\,cm$^{-2}$\,s$^{-1}$ (Inoue et al.\ 1981),
but peak fluxes down to about 2.0$\times$10$^{-8}$\,erg\,cm$^{-2}$\,s$^{-1}$ were reported 
(Makishima et al.\ 1981). During a recent outburst fifteen X-ray bursts were seen by the RXTE/PCA (Markwardt et al.\ 2000).
No photospheric radius expansion and only modest cooling were reported.

We find that two of the X-ray bursts seen by the RXTE/PCA 
were much brighter than the other thirteen. Our analysis shows (for the first time) that they both have a weak 
(factor of $\sim$2) photospheric radius expansion phase and 
clear cooling during the decay (see Fig.~\ref{plot_5RXTEbursts}). Maximum bolometric fluxes 
($\simeq$4.75$\times$10$^{-8}$\,erg\,cm$^{-2}$\,s$^{-1}$) are reached near touch-down.
The weaker X-ray bursts did not show evidence for photospheric radius expansion.

\subsection{MX\,1746$-$20/NGC\,6440}

Observations with the BeppoSAX/WFC and NFI of the transient MX\,1746$-$20 
during its second ever recorded outburst revealed four X-ray bursts (in~'t~Zand et al.\ 1999). 
The bolometric peak flux during the first 12\,s interval of the $\sim$100\,s long X-ray burst seen with the NFI was 
1.77$\pm$0.14$\times$10$^{-8}$\,erg\,cm$^{-2}$\,s$^{-1}$. No evidence for photospheric radius expansion
was seen.

\subsection{4U\,1746$-$37/NGC\,6441}

Various X-ray bursts have been seen from the location of 4U\,1746$-$37, i.e.\ by
SAS-3/HTC (Li \&\ Clark 1977; Cominsky 1981), EXOSAT/ME (Sztajno et al.\ 1987), Ginga/LAC (Sansom et al.\ 1993),
BeppoSAX/NFI (Parmar et al.\ 1999) and RXTE/PCA (Jonker et al.\ 2000). One of the two X-ray bursts seen with 
the SAS-3/HTC had a peak bolometric luminosity of 0.50$\pm$0.17$\times$10$^{-8}$\,erg\,cm$^{-2}$\,s$^{-1}$
(Cominksy 1981). The two X-ray bursts seen with EXOSAT/ME both showed weak (factor $\sim$3) photospheric radius expansion. 
Their peak fluxes were 1.0$\pm$0.1 and 0.9$\pm$0.1$\times$10$^{-8}$\,erg\,cm$^{-2}$\,s$^{-1}$
(Sztajno et al.\ 1987). Sztajno et al.\ (1987) also performed a re-analysis of the two SAS-3/HTC X-ray bursts.
They found 0.75$\pm$0.26 and 0.46$\pm$0.23$\times$10$^{-8}$\,erg\,cm$^{-2}$\,s$^{-1}$, respectively.

Eight of the twelve X-ray bursts reported by Jonker et al.\ (2000) were bright. Our analysis 
shows that none of the X-ray bursts showed photospheric radius expansion. An example of the fit parameter results of one
of the eight bright X-ray bursts is given in Fig.~\ref{plot_5RXTEbursts}. 
The eight bright X-ray bursts reached bolometric peak fluxes in the range 
$\simeq$0.4--0.6$\times$10$^{-8}$\,erg\,cm$^{-2}$\,s$^{-1}$.

\subsection{GRS\,1747$-$312/Terzan 6}

X-ray bursts from the transient source GRS\,1747$-$312 were only recently found during an outburst
(in~'t~Zand et al.\ 2002, in preparation). 
One of these showed photospheric radius expansion of a factor of $\sim$5 with a bolometric peak flux of 
2.10$\pm$0.07$\times$10$^{-8}$\,erg\,cm$^{-2}$\,s$^{-1}$.

\subsection{4U\,1820$-$30/NGC\,6624}

4U\,1820$-$30 was the first source seen to exhibit X-ray bursts (Grindlay et al.\ 1976). One of these two 
X-ray bursts seen by the ANS/HXX had a peak flux of about 6.5$\times$10$^{-8}$\,erg\,cm$^{-2}$\,s$^{-1}$ 
(1--15\,keV; re-analysis by Grindlay,
private communication in Vacca et al.\ 1986). The 22 X-ray bursts seen by SAS-3/RMC (Clark et al.\ 1977) 
had a mean peak flux of 6.0$\times$10$^{-8}$\,erg\,cm$^{-2}$\,s$^{-1}$ (re-analysis by G.W.~Clark,
private communication in Vacca et al.\ 1986). A subset (five) of these X-ray bursts (Clark et al.\ 1976)
showed a mean peak flux of 5.2$\times$10$^{-8}$\,erg\,cm$^{-2}$\,s$^{-1}$ (re-analysis by G.W.~Clark,
private communication in Vacca et al.\ 1986). Another subset (six) of the SAS-3/RMC X-ray bursts were analysed
in more detail by Vacca et al.\ (1986). They concluded that the bolometric peak fluxes, which ranged from 
3.9$\pm$0.2 to 5.3$\pm$0.4$\times$10$^{-8}$\,erg\,cm$^{-2}$\,s$^{-1}$, were the same within the errors.
The weighted mean peak value was 4.2$\times$10$^{-8}$\,erg\,cm$^{-2}$\,s$^{-1}$, with an estimated uncertainty 
of about 10\%, allowing for systematic errors.
All six X-ray bursts showed photospheric radius expansion. 
EXOSAT/ME saw seven radius expansion bursts (Haberl et al.\ 1987). 
Individual flux measurements (0.25\,s or 0.5\,s resolution) during the photospheric radius expansion phase 
($\sim$3\,s duration) ranged from 3 to 8$\times$10$^{-8}$\,erg\,cm$^{-2}$\,s$^{-1}$, whereas the 
average peak luminosity during this first 3\,s ranged from 4.1 to 5.7$\times$10$^{-8}$\,erg\,cm$^{-2}$\,s$^{-1}$,
with a mean of 5.1$\times$10$^{-8}$\,erg\,cm$^{-2}$\,s$^{-1}$. (Note that van Paradijs \&\ Lewin (1987)
used the results presented by Haberl et al.\ (1987) to derive a maximum flux during the photospheric radius expansion
phase of 5.28$\pm$0.19$\times$10$^{-8}$\,erg\,cm$^{-2}$\,s$^{-1}$.)
A re-analysis of the EXOSAT/ME X-ray bursts 
showed that the bolometric flux at touch down ranges between 4.3 and 5.0$\times$10$^{-8}$\,erg\,cm$^{-2}$\,s$^{-1}$,
with a mean of 4.65$\times$10$^{-8}$\,erg\,cm$^{-2}$\,s$^{-1}$
(Damen et al.\ 1990). More recently, the RXTE/PCA observed a short ($\sim$25\,s) X-ray burst (Zhang et al.\ 1998), 
which was also a radius expansion burst (Franco \&\ Strohmayer 1999; see also Strohmayer \&\ Brown 2002). 
Another very long ($>$2.5\,hr) X-ray burst (so-called superburst), also seen with RXTE/PCA, showed
strong photospheric radius expansion and reached a bolometric peak flux of 
6.5$\times$10$^{-8}$\,erg\,cm$^{-2}$\,s$^{-1}$ (Strohmayer \&\ Brown 2002).

Our analysis of the normal radius expansion burst seen by RXTE/PCA showed indeed 
strong photospheric radius expansion, by a factor of $\sim$20, see Fig.~\ref{plot_5RXTEbursts}. 
Note the strong similarity with the X-ray burst 
seen from XB\,1745$-$25, except for the brief stronger expansion in 4U\,1820$-$30
(Fig.~\ref{plot_5RXTEbursts}).
The BeppoSAX/WFCs have seen many X-ray X-ray bursts from 4U\,1820$-$30
(see. e.g.\ Fig.~4). From our analysis of 
21 X-ray X-ray bursts we identified 15 radius expansion bursts (see e.g.\ Fig.~5). 
The maximum peak fluxes of all the radius expansion bursts do reach 
the same value of 5.9$\pm$0.3$\times$10$^{-8}$\,erg\,cm$^{-2}$\,s$^{-1}$
with a $\chi^2_{\rm red}$ of 0.82 for 14 dof.

\subsection{H1825$-$331/NGC\,6652}

Two type~I X-ray bursts were first observed by the BeppoSAX/WFC from the transient H1825$-$331
(in~'t~Zand et al.\ 1998). A bolometric peak flux of about 0.8$\times$10$^{-8}$\,erg\,cm$^{-2}$\,s$^{-1}$
was found over the first 16\,s of each X-ray burst, which had exponential decay times of about 16\,s and
27\,s (2--8\,keV). During a serendipitous ASCA observation Mukai \&\ Smale (2000) found another
X-ray burst. It reached a peak flux of about 0.2$\times$10$^{-8}$\,erg\,cm$^{-2}$\,s$^{-1}$
during the first 30\,s of the X-ray burst. No photospheric radius expansion was seen in both cases.

The RXTE/PCA observed a bright and a weak X-ray burst from H1825$-$331. No clear evidence for photospheric radius expansion is found, 
but we note that near maximum of the bright X-ray burst the black-body 
temperature shows a short ($\sim$1\,s) 
and small dip, while the apparent black-body radius shows a small hump
(Fig.~\ref{plot_pars_terzan2_ngc7078_ngc6652}). This might indicate that the 
event reached the Eddington limit, but that it was not strong enough to 
cause the photosphere to expand appreciably.

\subsection{A1850$-$08/NGC\,6712}

An X-ray burst with a peak intensity of about 1.5~Crab was observed from the vicinity of A1850$-$08 with OSO-8
(Swank et al.\ 1976). We note here that the peak intensity of the long X-ray burst from 4U\,1722$-$30
observed by OSO-8 (see Sect.~B.2) also reached 1.5~Crab (Swank et al.\ 1977). Assuming the X-ray burst spectra are similar during
the peak (which may not necessarily be the case since the event from 4U\,1722$-$30 showed photospheric radius
expansion), one may infer a peak flux of about 6$\times$10$^{-8}$\,erg\,cm$^{-2}$\,s$^{-1}$.
An X-ray flare was reported by Cominsky et al.\ (1977) using data from Uhuru.
It increased by a factor of 40 in less than 5.5\,min. The next 16.7\,min the intensity decreased with
an exponential decay time of 3.1$\pm$0.4\,min. It may have persisted for up to 78\,min.
The observed peak flux was about 0.6$\times$10$^{-8}$\,erg\,cm$^{-2}$\,s$^{-1}$. 
Three X-ray bursts were seen with SAS-3/HTC, of which one showed a weak (factor $\sim$2) 
photospheric radius expansion (Hoffman et al.\ 1980; see also Cominsky 1981). Hoffman et al.\ (1980) found a 
bolometric peak flux of 6.0$\pm$0.2$\times$10$^{-8}$\,erg\,cm$^{-2}$\,s$^{-1}$, whereas 
Cominsky (1981) report 5.2$\pm$0.5$\times$10$^{-8}$\,erg\,cm$^{-2}$\,s$^{-1}$ for the same X-ray burst. 
The bolometric peak flux we use in our investigation, as indicated with a
$\surd$ in Table~\ref{account}, is the value derived by Cominsky (1981). 
The fluxes quoted by the Hoffman et al.\ (1980) are
calculated from conversions from count rates to photon rates using a Crab-like spectrum, whereas 
those quoted by the former derive conversions using black-body spectra (see Cominsky 1981 for
a more detailed discussion).

\subsection{4U\,2129+12/NGC\,7078}

A very strong (factor $\gtrsim$100) radius expansion burst was observed by Ginga/LAC from
4U\,2129+12 (Dotani et al.\ 1990;
van Paradijs et al.\ 1990). The bolometric peak flux was about 4.2$\times$10$^{-8}$\,erg\,cm$^{-2}$\,s$^{-1}$,
with a conservative error of 0.1$\times$10$^{-8}$\,erg\,cm$^{-2}$\,s$^{-1}$.
A less strong (factor $\sim$10) radius expansion burst was recently seen by RXTE/PCA, which reached a 
bolometric peak flux of 5$\times$10$^{-8}$\,erg\,cm$^{-2}$\,s$^{-1}$ (Smale 2001). Many more X-ray bursts were identified
in the RXTE/ASM data (Charles et al.\ 2002). 
A Chandra ACIS-S/HETG observation revealed that 4U\,2129+12 is in fact composed of two
bright sources, separated by only 2".7 (White \&\ Angelini 2001). 
One source, NGC\,7078 X-1, is associated with AC~211, the other is identified as NGC\,7078 X-2.
NGC\,7078 X-2 is the most likely source for the X-ray bursts.

Our reanalysis of the event seen by RXTE/PCA showed similar results (Fig.~\ref{plot_pars_terzan2_ngc7078_ngc6652}) 
as those reported by Smale (2001). 
We found a bolometric peak flux of $\sim$4$\times$10$^{-8}$\,erg\,cm$^{-2}$\,s$^{-1}$ near touch-down.
The BeppoSAX/WFC also observed one X-ray burst (Figs.~\ref{plot_WFC_lc} and \ref{plot_pars_WFC}). However, no evidence
for photospheric radius expansion was found, although it reached a comparable bolometric peak flux 
($\sim$4$\times$10$^{-8}$\,erg\,cm$^{-2}$\,s$^{-1}$) as seen during the
strong radius expansion burst observed by Ginga/LAC and RXTE/PCA.
Note that our method of subtracting the persistent pre-burst emission automatically also removes any contribution 
of AC~211/NGC\,7078 X-1.

\begin{table*}
\caption{Observed peak fluxes for the
twelve globular cluster X-ray bursters. Peak fluxes from RXTE/PCA observations derived in this
paper have been corrected, see Sect.~3.5; the uncorrected values are given between brackets.
Also given are the instrument with which the events were seen (BSAX refers to BeppoSAX),
whether photospheric radius expansion (RE) was reported or not, and comments. 
X-ray bursts indicated with a $\surd$ are used to derive the average weighted 
bolometric peak flux of radius expansion bursts, 
$\overline{F_{\rm bb,peak}}$, or the maximum observed bolometric peak flux for normal X-ray bursts, 
$F_{\rm bb,max}$, see Table 2.
References to the relevant paper are given in brackets in the sixth column:
[1] Forman \&\ Jones (1976),
[2] Cominsky (1981),
[3] this paper,
[4] Swank et al.\ (1977),
[5] Grindlay et al.\ (1980),
[6] Guainazzi et al.\ (1998),
[7] Molkov et al.\ (2000),
[8] Marshall et al.\ (1979),
[9] Inoue et al.\ (1981),
[10] Pavlinsky et al.\ (2001),
[11] Makishima et al.\ (1981),
[12] in~'t~Zand et al.\ (1999),
[13] Sztajno et al.\ (1987),
[14] in~'t~Zand et al.\ (2002, in preparation),
[15] Vacca et al.\ (1986),
[16] van Paradijs \&\ Lewin (1987),
[17] Damen et al.\ (1990),
[18] Strohmayer \&\ Brown (2002),
[19] in~'t~Zand et al.\ (1998),
[20] Mukai \&\ Smale (2000),
[21] Swank et al.\ (1976),
[22] Cominsky et al.\ (1977),
[23] Hoffman et al.\ (1980),
[24] van Paradijs et al.\ (1990),
[25] Smale (2001).
}
\begin{tabular}{ccccccl}
\hline
\multicolumn{7}{c}{~} \\
\multicolumn{1}{c}{X-ray burster} & \multicolumn{1}{c}{peak flux} &
\multicolumn{1}{c}{satellite/} & \multicolumn{1}{c}{RE?} &
\multicolumn{1}{c}{$\overline{F_{\rm bb,peak}}$} &
\multicolumn{1}{c}{ref} &
\multicolumn{1}{l}{\#\ of bursts + comment} \\
\multicolumn{1}{c}{~} & \multicolumn{1}{c}{ (10$^{-8}$\,erg\,cm$^{-2}$\,s$^{-1}$)} &
\multicolumn{1}{c}{instrument} & \multicolumn{1}{c}{~} &
\multicolumn{1}{c}{or $F_{\rm bb,max}$} &
\multicolumn{1}{c}{~} &
\multicolumn{1}{c}{~} \\
\multicolumn{7}{c}{~} \\
\hline
MX\,0513$-$40   & 0.9--1.7    & SAS-3/HTC    & no  & & [2] & range of 4 \\
                & 1.26$\pm$0.05 (1.52$\pm$0.06) & RXTE/PCA   & no  & & [3] & 1  \\
                & 1.99$\pm$0.20 & BSAX/WFC & yes & $\surd$ & [3] & 1; strong RE \\
                & 2.05$\pm$0.39 & BSAX/WFC & yes & $\surd$ & [3] & 1; weak RE \\
4U\,1722$-$30   & $\simeq$6.2 & OSO-8        & yes & & [4] & 1; strong RE \\
                & $\simeq$5.2 & Einstein/MPC & yes & & [5] & 1; weak RE \\
                & $\simeq$5.0 & BSAX/MECS & no  & & [6] & 1 \\
                & 6.23$\pm$0.15 & BSAX/WFC & yes & $\surd$ & [3] & 24; all strong RE \\
                & 7.29$\pm$0.06 & RXTE/PCA   & yes & & [7] & 1; strong RE \\
                & 6.11$\pm$0.10 (7.59$\pm$0.12) & RXTE/PCA   & yes & $\surd$ & [3] & 1 (same as above); strong RE \\
MXB\,1730$-$335 & $\simeq$1.7 & SAS-3/HTC    & no  & & [8] & average of many X-ray bursts \\
                & 0.9--1.7 (1.1--2.1) & RXTE/PCA     & no  & & [3] & 4\\
                & 1.66$\pm$0.08 (2.06$\pm$0.10) & RXTE/PCA   & no  & $\surd$ & [3] & 1 (of the above 4) \\
XB\,1733$-$30   & 7.4$\pm$1.0 & Hakucho/FMC  & no  & $\surd$ & [9] & 1 \\
                & 6.3$\pm$0.8 & Granat/ART-P & no  & & [10] & 1 \\
XB\,1745$-$25   & 2.0--6.1    & Hakucho/CMC/FMC & no  & & [11] & 14 \\
                & 6.1$\pm$0.7 & Hakucho/FMC  & no  & & [9] & 1 (of the above 14) \\
                & 4.74$\pm$0.18 (5.89$\pm$0.22) & RXTE/PCA   & yes & $\surd$ & [3] & 1; weak RE \\
                & 4.75$\pm$0.17 (5.90$\pm$0.21) & RXTE/PCA   & yes & $\surd$ & [3] & 1; weak RE \\
MX\,1746$-$20   & 1.77$\pm$0.14 & BSAX/NFI & no & $\surd$ & [12] & 1 \\
4U\,1746$-$37   & 0.50$\pm$0.17 & SAS-3/HTC  & no  & & [2] & 1 \\
                & 0.75$\pm$0.27 & SAS-3/HTC  & no  & & [13] & 1 (same as above) \\
                & 0.46$\pm$0.23 & SAS-3/HTC  & no  & & [13] & 1 \\
                & 1.0$\pm$0.1 & EXOSAT/ME    & yes & $\surd$ & [13] & 1; weak RE \\
                & 0.9$\pm$0.1 & EXOSAT/ME    & yes & $\surd$ & [13] & 1; weak RE \\
                & 0.4--0.6 (0.5--0.75) & RXTE/PCA     & no  & & [3] & 8 \\
GRS\,1747$-$312 & 1.71$\pm$0.06 & RXTE/PCA   & yes & $\surd$ & [14] & 1; weak RE \\
4U\,1820$-$30   & 6.5$\pm$1.3 & ANS/HXX      & no  & & [15] & 1 \\
                & $\simeq$6.0 & SAS-3/RMC    & no  & & [15] & 22 \\
                & $\simeq$5.2 & SAS-3/RMC    & no  & & [15] & 5 (of the above 22) \\
                & 4.2$\pm$0.4 & SAS-3/RMC    & yes & $\surd$ & [15] & 6 (of the above 22); all strong RE \\
                & 5.28$\pm$0.19 & EXOSAT/ME  & yes & $\surd$ & [16] & 7; all strong RE \\
                & $\simeq$4.65 & EXOSAT/ME   & yes & & [17] & 7 (same as above); all strong RE \\
                & $\simeq$6.5 & RXTE/PCA     & yes & & [18] & 1 (superburst); strong RE \\
                & 5.26$\pm$0.12 (6.53$\pm$0.15) & RXTE/PCA   & yes & $\surd$ & [3] & 1; strong RE \\
                & 5.9$\pm$0.3 & BSAX/WFC & yes & $\surd$ & [3] & 15; all strong RE \\
H1825$-$331     & $\simeq$0.8 & BSAX/WFC & no  & & [19] & 2 \\
                & $\simeq$0.2 & ASCA/SIS     & no  & & [20] & 1 \\
                & 2.87$\pm$0.07 (3.57$\pm$0.09) & RXTE/PCA   & no  & $\surd$ & [3] & 1 \\
A1850$-$08      & $\simeq$6   & OSO-8        & no  & & [21] & 1 \\
                & 6.0$\pm$0.2 & SAS-3/HTC    & yes & & [23] & 1; weak RE \\
                & 5.2$\pm$0.5 & SAS-3/HTC    & yes & $\surd$ & [2] & 1; weak RE (same as above) \\
4U\,2129+12     & 4.2$\pm$0.1 & Ginga/LAC    & yes & $\surd$ & [24] & 1; very strong RE \\
                & $\simeq$5.0 & RXTE/PCA     & yes & & [25] & 1; strong RE \\
                & 3.81$\pm$0.07 (4.68$\pm$0.09) & RXTE/PCA   & yes & $\surd$ & [3] & 1 (same as above); strong RE \\
                & 3.9$\pm$0.8 & BSAX/WFC & no  & & [3] & 1\\
\hline
\end{tabular}
\label{allbursts}
\end{table*}


\begin{thebibliography}{}
\bibitem[2000]{}
Barret, D., Olive, J.F., Boirin, L., Done, C., Skinner, G.K., Grindlay, J.E.
2000, ApJ, 533, 329
\bibitem[1984]{}
Basinska, E.M., Lewin, W.H.G., Sztajno, M., Cominsky, L.R., Marshall, F.J. 1984, ApJ, 281, 337
\bibitem[1995]{}
Bildsten, L. 1995, ApJ, 438, 852
\bibitem[1998]{}
Bildsten, L. 1998, in: The Many Faces of Neutron Stars, R.~Buccheri, J.~van Paradijs,
M.A.~Alpar (eds.), Dordrecht: Kluwer, p.~419
\bibitem[1993]{}
Bradt, H.V., Rothschild, R.E., Swank, J.H. 1993, A\&AS, 97, 355
\bibitem[1998]{}
Chaboyer, B., Demarque, P., Kernan, P.J., Krauss, L.M. 1998, ApJ, 494, 96
\bibitem[2000]{}
Chaboyer, B., Sarajedini, A., Armandroff, T.E. 2000, AJ, 120, 3102
\bibitem[2002]{}
Charles, P.A., Clarkson, W.I., van Zyl, L. 2002, NewA, 7, 21
\bibitem[2001]{}
Chou, Y., Grindlay, J.E. 2001, ApJ, 563, 934
\bibitem[1997]{}
Christian, D.J., Swank, J.H. 1997, ApJS, 109, 177
\bibitem[1976]{}
Clark, G.W., Jernigan, J.G., Bradt, H., et al. 1976, ApJ, 207, L105
\bibitem[1977]{}
Clark, G.W., Li, F.K., Canizares, C., Hayakawa, S., Jernigan, G., Lewin, W.H.G. 1977, MNRAS, 179, 651
\bibitem[2000a]{}
Cocchi, M., Bazzano, A., Natalucci, L., Ubertini, P., Heise, J., Kuulkers, E., Muller, J.M., in~'t~Zand, J.J.M. 2000a,
A\&A, 357, 527
\bibitem[2000b]{}
Cocchi, M., Bazzano, A., Natalucci, L., Ubertini, P., Heise, J., Kuulkers, E., in~'t~Zand, J.J.M. 2000b,
in: The Fifth Compton Symposium, M.L.~McConnell, J.M.~Ryan (eds.), AIP Conf.\ Proc.\ 510, p.~217
\bibitem[2001]{}
Cocchi, M., Bazzano, A., Natalucci, L., Ubertini, P., Cornelisse, R., 
in~'t~Zand, J.J.M., Heise, J., Kuulkers, E., Smith, M.J.S. 2001, in Exploring the Gamma-Ray Universe,
Fourth INTEGRAL workshop, ESA SP-459, p.~279
\bibitem[2002]{}
Cohn, H.N., Lugger, P.M., Grindlay, J.E., Edmonds, P.D., 2002, ApJ, 571, 818
\bibitem[1981]{}
Cominsky, L.R. 1981, PhD thesis, MIT
\bibitem[1984]{}
Cominsky, L.R., Wood, K.S. 1984, ApJ, 283, 765
\bibitem[1977]{}
Cominsky, L., Forman, W., Jones, C., Tananbaum, H. 1977, ApJ, 211, L9
\bibitem[1990]{}
Damen, E., Magnier, E., Lewin, W.H.G., Tan, J., Penninx, W., van Paradijs, J. 1990, A\&A, 237, 103.
\bibitem[2000]{}
Davidge, T.J. 2000, ApJS, 126, 105
\bibitem[2000]{}
Deutsch, E.W., Margon, B., Anderson, S.F. 2000, ApJ, 520, L21
\bibitem[1990]{}
Dotani, T., Inoue, H., Murakami, T., et al. 1990, Nat, 347, 534
\bibitem[1976]{}
Forman, W., Jones, C. 1976, ApJ, 207, L177.
\bibitem[2001]{}
Franco, L.M. 2001, ApJ, 554, 340
\bibitem[1999]{}
Franco, L.M., Strohmayer, T.E. 1999, AAS, 195, $\#$126.09
\bibitem[1981]{}
Fujimoto, M.Y., Hanawa, T., Miyaji, S. 1981, ApJ, 247, 267
\bibitem[2002]{}
Galloway, D.K., Psaltis, D., Chakrabarty, D., Muno, M.P. 2002b, ApJ, submitted [astro-ph/0208464]
\bibitem[2002]{}
Galloway, D., Savov, P., Psaltis, D., Chakrabarty, D., Muno, M. 2002a, Bull.\ of the American Phyisical
Society, 47 (2), 180
\bibitem[1979]{}
Goldman, I. 1979, A\&A, 78, L15
\bibitem[1986]{}
Gottwald, M., Haberl, F., Parmar, A.N., White, N.E. 1986, ApJ, 308, 213
\bibitem[1976]{}
Grindlay, J., Gursky, H., Schnopper, H., Parsignault, D.R., Heise, J., Brinkman, A.C., 
Schrijver, J. 1976, ApJ, 205, L127
\bibitem[1980]{}
Grindlay, J.E., Marshall, H.L., Hertz, P., et al. 1980, ApJ, 240, L121
\bibitem[1999]{}
Guainazzi, M., Parmar, A.N., Oosterbroek, T. 1999, A\&A, 349, 819
\bibitem[1998]{}
Guainazzi, M., Parmar, A.N., Segreto, A., Stella, L., Dal Fiume, D., Oosterbroek, T. 1998, A\&A, 339, 802
\bibitem[1999]{}
Guerriero, R., Fox, D.W., Kommers, J., et al. 1999, MNRAS, 307, 179
\bibitem[1987]{}
Haberl, F., Stella, L., White, N.E., Priedhorsky, W.C., Gottwald, M. 1987, ApJ, 314, 266
\bibitem[1996]{}
Harris, W.E. 1996, AJ, 112, 1487
\bibitem[2000]{}
Heasley, J.N., Janes, K.A., Zinn, R., Demarque, P., Da Costa, G.S., Christian, C.A. 2000, AJ, 120, 879
\bibitem[1980]{}
Heinke, C.O., Edmonds, P.D., Grindlay, J.E. 2001, ApJ, 562, 363
\bibitem[1980]{}
Hoffman, J.A., Cominsky, L., Lewin, W.H.G. 1980, ApJ, 240, L27
\bibitem[1978]{}
Hoffman, J.A., Marshall, H.L., Lewin, W.H.G. 1978, Nat, 271, 630
\bibitem[2001]{}
Homer, L., Anderson, S.F., Margon, B., Deutsch, E.W., Downes, R.A. 2001, ApJ, 550, L155
\bibitem[1996]{}
Homer, L., Charles, P.A., Naylor, T., van Paradijs, J., Auri\`ere, M., Koch-Miramond, L. 1996, MNRAS, 282, L37
\bibitem[1992]{}
Hut, P., McMillan, S., Goodman, J., et al. 1992, PASP, 104, 981
\bibitem[1993]{}
Idiart, T.P., Barbuy, B., Perrin, M.-N., Ortolani, S., Bica, E., Renzini, A. 2002, A\&A, 381, 472
\bibitem[1993]{}
Ilovaisky, S.A., Auri\`ere, M., Koch-Miramond, L., et al. 1993, A\&A, 270, 139
\bibitem[1981]{}
Inoue, H., Koyama, K., Makishima, K., et al. 1981, ApJ, 250, L71
\bibitem[1984]{}
Inoue, H., Waki, I., Koyama, K., Matsuoka, M., Ohashi, T., Tanaka, Y., Tsunemi, H. 1984, PASJ, 36, 831
\bibitem[1999]{}
in 't Zand, J. 1999, Calibration report of the Wide Field Cameras onboard BeppoSAX
\bibitem[2001]{}
in 't Zand, J. 2001, in Exploring the Gamma-Ray Universe,
Fourth INTEGRAL workshop, ESA SP-459, p.~463
\bibitem[2000]{}
in 't Zand, J.J.M., Bazzano, A., Cocchi, M., et al. 2000, A\&A, 355, 145
\bibitem[2001]{}
in 't Zand, J.J.M., Cornelisse, R., Kuulkers, E., et al. 2001a, A\&A, 372, 916
\bibitem[1998]{}
in 't Zand, J.J.M., Verbunt, F., Heise, J., Muller, J.M., Bazzano, A., Cocchi, M., 
Natalucci, L., Ubertini, P. 1998, A\&A, 329, L37
\bibitem[1999]{}
in 't Zand, J.J.M., Verbunt, F., Strohmayer, T.E., et al. 1999, A\&A, 345, 100
\bibitem[2001]{}
in 't Zand, J.J.M., van Kerkwijk, M.H., Pooley, D., Verbunt, F., Wijnands, R., Lewin, W.H.G. 2001b,
ApJ, 563, L41
\bibitem[1997]{}
Jager, R., Mels, W.A., Brinkman, A.C., et al. 1997, A\&ASS, 125, 557
\bibitem[2000]{}
Jahoda, K. 2000, talk presented at the meeting 
``Rossi 2000: Astrophysics with the Rossi X-ray Timing Explorer'', Greenbelt, Maryland, USA;
see {\tt http://lheawww.gsfc.nasa.gov/users/keith/ rossi2000/energy\_response.ps}
\bibitem[2001]{}
Jonker, P.G., van der Klis, M., Homan, J., Wijnands, R., van Paradijs, J., M\'endez, M., Kuulkers, E., 
Ford, E.C. 2000, ApJ, 531, 453
\bibitem[2001]{}
Juett, A.M., Psaltis, D., Chakrabarty, D. 2001, ApJ, 560, L59
\bibitem[1997]{}
Kommers, J.M., Fox, D.W., Lewin, W.H.G., Rutledge, R.E., van Paradijs, J., Kouveliotou, C. 1997,
ApJ, 482, L53
\bibitem[1999]{}
Kuiper, L., Hermsen, W., Krijger, J.M., Bennett, K., Carrami\~nana, Sch\"onfelder, V., Bailes, M., 
Manchester, R.N. 1999, A\&A, 351, 199
\bibitem[2002]{}
Kuulkers, E., Homan, J., van der Klis, M., Lewin, W.H.G., M\'endez, M. 2002, A\&A, 382, 947
\bibitem[1982]{}
Lewin, W.H.G. 1982, in: Accreting Neutron Stars, W.~Brinkmann \&\ J.~Tr\"umper (eds.),
MPE Report 177, ISSN 0340-8922, p.~76
\bibitem[1985]{}
Lewin, W.H.G. 1985, in: Galactic and Extragalactic Compact X-ray Sources,
Y.~Tanaka \&\ W.H.G.~Lewin (eds.), Institute of Space and Astronautical Science, Tokyo, p.~89
\bibitem[1996]{}
Lewin, W.H.G., Rutledge, R.E., Kommers, J.M., van Paradijs, J., Kouveliotou, C. 1996,
ApJ, 462, L39
\bibitem[1984]{}
Lewin, W.H.G., Vacca, W.D., Basinska, E.M. 1984, ApJ, 277, L57
\bibitem[1993]{}
Lewin, W.H.G., van Paradijs, J., Taam, R.E. 1993, Space Sci.\ Rev., 62, 223
\bibitem[1977]{}
Li, F., Clark, G. 1977, IAU Circ.\ 3095
\bibitem[1984]{}
London, R.A., Taam, R.E., Howard, W.M. 1984, ApJ, 287, L27
\bibitem[1986]{}
London, R.A., Taam, R.E., Howard, W.M. 1986, ApJ, 306, 170
\bibitem[1997]{}
Madej, J. 1997, A\&A, 320, 177
\bibitem[1981]{}
Makishima, K., Ohashi, T., Inoue, H., et al. 1981, ApJ, 247, L23
\bibitem[2000]{}
Markwardt, C.B., Strohmayer, T.E., Swank, J.H., Zhang, W. 2000, IAUC, 7482
\bibitem[1979]{}
Marshall, H.L., Hoffman, J.A., Doty, J., Lewin, W.H.G., Ulmer, M.P. 1979, ApJ, 227, 555
\bibitem[2000]{}
Masetti, N., Frontera, F., Stella, L., et al. 2000, A\&A, 363, 188
\bibitem[1984]{}
Melia, F. 1987, ApJ, 315, L43
\bibitem[1984]{}
Melia, F., Joss, P.C. 1985, ApJ, 295, 98
\bibitem[2000]{}
Molkov, S.V., Grebenev, S.A., Luovinov, A.A. 2000, A\&A, 357, L41
\bibitem[1983]{}
Morrison, R., McCammon, D. 1983, ApJ, 270, 119
\bibitem[2000]{}
Mukai, K., Smale, A.P. 2000, ApJ, 533, 352
\bibitem[2000]{}
Muno, M.P., Fox, D.W., Morgan, E.H., Bildsten, L. 2000, ApJ, 542, 1016
\bibitem[1986]{}
Nelson. L.A., Rappaport, S.A., Joss, P.C. 1986, ApJ, 304, 231
\bibitem[2002]{}
Origlia, L., Rich, R.M., Castro, S. 2002, AJ, 123, 1559
\bibitem[1999]{}
Orosz, J.A., Kuulkers, E. 1999, MNRAS, 305, 132
\bibitem[2001]{}
Paltrinieri, B., Ferraro, F.R., Paresce, F., De Marchi, G. 2001, AJ, 121, 3114
\bibitem[1985]{}
Parmar, A.N., Smith, A. 1985, EXOSAT Express, No.~10, p.~40
\bibitem[1999]{}
Parmar, A.N., Oosterbroek, T., Guainazzi, M., Segreto, A., dal Fiume, D., Stella, L. 1999, A\&A, 351, 225
\bibitem[2001]{}
Parmar, A.N., Oosterbroek, T., Sidoli, L., Stella, L., Frontera, F. 2001, A\&A, 380, 490
\bibitem[1989]{}
Parmar, A.N., Stella, L., Giommi, P. 1989, A\&A, 222, 96
\bibitem[1986]{}
Parmar, A.N., White, N.E., Giommi, P., Gottwald, M. 1986, ApJ, 308, 199
\bibitem[2001]{}
Pavlinsky, M.N., Grebenev, S.A., Lutovinov, A.A., Sunyaev, R.A., Finoguenov, A.V. 2001,
Astronomy Letters, 27, 297
\bibitem[1993]{}
Predehl, P., Schmitt, J.H.M.M. 1995, A\&A, 293, 889
\bibitem[2001]{}
Pritzl, B.J., Smith, H.A., Catelan, M., Sweigart, A.V. 2001, AJ, 122, 2600
\bibitem[1993]{}
Sansom, A.E., Dotani, T., Asai, K., Lehto, H.J. 1993, MNRAS, 262, 429
\bibitem[1986]{}
Schattenburg, M.L., Canizares, M.L. 1986, ApJ, 301, 759
\bibitem[2001]{}
Sidoli, L., Parmar, A.N., Oosterbroek, T., Stella, L., Verbunt, F., Masetti, N., 
Dal Fiume, D. 2001, A\&A, 368, 451
\bibitem[2001]{}
Smale, A.P. 2001, ApJ, 562, 957
\bibitem[2001]{}
Stella, L. 2001, in X-ray Astronomy --- Stellar Endpoints, AGN, and the Diffuse X-ray
Background, ed.\ N.E.~White, G.~Malaguti, \&\ G.G.C.~Palumbo,
AIP Conf.\ Proc.\ Vol.~599, p.~365
\bibitem[1987]{}
Stella, L., Priedhorsky, W., White, N.E. 1987, ApJ, 312, L17
\bibitem[1999]{}
Stella, L., Vietri, M., Morsink, M. 1999, ApJ, 524, L63
\bibitem[1985]{}
Stollman, G.M., van Paradijs, J. 1985, A\&A, 153, 99
\bibitem[2002]{}
Strohmayer, T.E., Brown, E.F. 2002, ApJ, 566, 1045
\bibitem[1984]{}
Sugimoto, D., Ebisuzaki, T., Hanawa, T. 1984, PASJ, 36, 839
\bibitem[1977]{}
Swank, J.H., Becker, R.H., Boldt, E.A., Holt, S.S., Pravdo, S.H., Serlemitsos, P.J. 1977, ApJ 212, L73
\bibitem[1976]{}
Swank, J.H., Becker, R.H., Pravdo, S.H., Saba, J.R., Serlemitsos, P.J. 1976, IAU Circ.\ 3010
\bibitem[1987]{}
Sztajno, M., Fujimoto, M.Y., van Paradijs, J., Vacca, W.D., Lewin, W.H.G., Penninx, W., 
Tr\"umper, J. 1987, MNRAS, 226, 39
\bibitem[1984]{}
Tawara, Y., Kii, T., Hayakawa, S., et al. 1984, ApJ, 276, L41
\bibitem[1999]{}
Titarchuk, L., 1994, ApJ, 434, 313
\bibitem[1999]{}
Tomsick, J.A., Kaaret, P., Kroeger, R.A., Remillard, R.A. 1999, ApJ, 512, 892
\bibitem[1974]{}
Toor, A., Seward, F.D. 1974, AJ, 79, 995
\bibitem[1989]{}
Turner, M.J.L., Thomas, H.D., Patchett, B.E., et al. 1989, PASJ, 41, 345
\bibitem[1986]{}
Vacca, W.D., Lewin, W.H.G., van Paradijs, J. 1986, MNRAS, 220, 339
\bibitem[1978]{}
van Paradijs, J. 1978, Nat, 274, 650
\bibitem[1979]{}
van Paradijs, J. 1979, ApJ, 234, 609
\bibitem[1981]{}
van Paradijs, J. 1981, A\&A, 101, 174
\bibitem[1982]{}
van Paradijs, J. 1982, A\&A, 107, 51
\bibitem[1987]{}
van Paradijs, J., Lewin, W.H.G. 1987, A\&A, 172, L20
\bibitem[1990]{}
van Paradijs, J., Dotani, T., Tanaka, Y., Tsuru, T. 1990, PASJ, 42, 633
\bibitem[2001]{}
van Straaten, S., van der Klis, M., Kuulkers, E., M\'endez, M. 2001, ApJ, 551, 907
\bibitem[1997]{}
Verbunt, F. 1987, ApJ, 312, L23
\bibitem[1984]{}
Verbunt, F., van Paradijs, J., Elson, R., 1984, MNRAS, 210, 899
\bibitem[1995]{}
Verbunt, F., Bunk, W., Hasinger, G., Johnston, H.M. 1995, A\&A, 300, 732
\bibitem[2001]{}
White, N.E., Angelini, L. 2001, ApJ, 561, L101
\bibitem[2002]{}
Wijnands, R., Muno, M.P., Miller, J.M., Franco, L.M., Strohmayer, T., Galloway, D., 
Chakrabarty, D., 2002, ApJ, 566, 1060
\bibitem[2001]{}
Wilms, J., Nowak, M.A., Dove, J.B., Fender, R.P., Di Matteo, T. 1999, ApJ, 522, 460
\bibitem[1998]{}
Zhang, W., Smale, A.P., Strohmayer, T.E., Swank, J.H. 1998, ApJ, 500, L171

\end{thebibliography}
\end{document}